\documentclass[aps,prc,twocolumn,showpacs,nofootinbib,superscriptaddress,groupedaddress]{revtex4}
\usepackage{dcolumn}   
\usepackage{bm}        
\usepackage{multirow}

\usepackage{amsmath}
\usepackage{amsfonts}
\usepackage{amssymb}
\usepackage{makeidx}
\usepackage{graphicx}
\usepackage{anysize}
\usepackage{hyperref}

\usepackage{slashed}

\newenvironment{mymathbox}
{\par\smallskip\centering\begin{lrbox}{0}%
\begin{minipage}[c]{0.8\textwidth}}
{\end{minipage}\end{lrbox}%
\framebox[0.9\textwidth]{\usebox{0}}%
\par\medskip
\ignorespacesafterend}
\newcommand{\bb}{\begin{mymathbox}}
\newcommand{\eb}{\end{mymathbox}}

\newcommand{\be}{\begin{equation}}
\newcommand{\ee}{\end{equation}}
\newcommand{\ba}{\begin{eqnarray}}
\newcommand{\ea}{\end{eqnarray}}
\newcommand{\fl}[1]{\begin{flalign}#1\end{flalign}}

\newcommand{\ubarN}{\overline{u}({\bf p}_N,s_N)}

\newcommand{\nk}{{\bf      k}}
\newcommand{\np}{{\bf      p}}
\newcommand{\nq}{{\bf      q}}
\newcommand{\nr}{{\bf      r}}

\newcommand{\nz}{{\bf      z}}

\newcommand{\npsi}{{\bf \npsi}}

\newcommand{\psib}{\overline{\psi}}

\newcommand{\phib}{\phi^*}

\newcommand{\de}{\text{d}}

\newcommand{\non}{\nonumber}

\newcommand{\bma}{\begin{pmatrix}}
\newcommand{\ema}{\end{pmatrix}}

\usepackage[dvips]{color}

\begin{document}

\title{Pion production within the hybrid relativistic plane wave impulse approximation model at MiniBooNE and MINERvA kinematics}

\author{R.~Gonz\'alez-Jim\'enez}
\affiliation{Department of Physics and Astronomy, Ghent University, Proeftuinstraat 86, B-9000 Gent, Belgium}
\author{K.~Niewczas}
\affiliation{Department of Physics and Astronomy, Ghent University, Proeftuinstraat 86, B-9000 Gent, Belgium}
\affiliation{Institute of Theoretical Physics, University of Wroc{\l}aw, Plac Maxa Borna 9, 50-204 Wroc{\l}aw, Poland}
\author{N.~Jachowicz}
\affiliation{Department of Physics and Astronomy, Ghent University, Proeftuinstraat 86, B-9000 Gent, Belgium}

\date{\today}

\begin{abstract}
The hybrid model for electroweak single-pion production (SPP) off the nucleon, presented in [Gonz\'alez-Jim\'enez et al., Phys. Rev. D 95, 113007 (2017)], is extended here to the case of incoherent pion-production on the nucleus. Combining a low-energy model with a Regge approach, this model provides valid predictions in the entire energy region of interest for current and future accelerator-based neutrino-oscillation experiments.
The Relativistic Mean-Field model is used for the description of the bound nucleons while the outgoing hadrons are considered as plane waves. This approach, known as Relativistic Plane-Wave Impulse Approximation (RPWIA), is a first step towards the development of more sophisticated models, it is also a test of our current understanding of the elementary reaction.
We focus on the charged-current $\nu$($\bar\nu$)-nucleus interaction at MiniBooNE and MINERvA kinematics. 
The effect on the cross sections of the final-state interactions, which affect the outgoing hadrons on their way out of the nucleus, is judged by comparing our results with those from the NuWro Monte Carlo event generator.
We find that the hybrid-RPWIA predictions largely underestimate the MiniBooNE data. In the case of MINERvA, our results fall below the $\nu$-induced 1$\pi^0$ production data, while a better agreement is found for $\nu$-induced 1$\pi^+$ and $\bar\nu$-induced 1$\pi^0$ production. 
\end{abstract}


\maketitle

\section{Introduction}\label{sec:intro}
 
Single-pion production (SPP) constitutes a significant contribution to neutrino-nucleus cross sections in the energy range covered by the neutrino experiments
K2K~\cite{K2K05}, MiniBooNE~\cite{MBNCpion010,MBCCpion011,MBCCpionC11}, MINERvA~\cite{MINERvACCpi15,MINERvACCpi16,MINERvAnuCC17}, SciBooNE~\cite{SciBooNE11}, T2K~\cite{T2Kinc13}, NOvA~\cite{NOvAweb}, and the future DUNE~\cite{DUNEweb} and HyperKamiokande~\cite{HyperK15}.
As the energy of the neutrino beam in future experiments (e.g.~DUNE) shifts to higher energies, the importance of the pion-production contribution, as compared to the quasielastic (QE) channel, increases.
Therefore, having theoretical models capable of providing accurate predictions of this reaction channel is essential to reduce the systematic uncertainties that plague the neutrino-oscillation analyses~\cite{Alvarez-Ruso17}.
In addition to that, the investigation of the neutrino-nucleon/nucleus interaction, beyond its role in the neutrino oscillation program, is of great interest itself, since it provides unique information on the weak response (axial-vector current) of nuclei and nucleons. This is important, for instance, in disentangling the electroweak structure of the nucleon and its resonances.

A variety of models describing neutrino-induced SPP are available in the literature~\cite{Rein81,Ahmad06,Hernandez07,Buss07,Praet09,Martini09,Serot12,Lalakulich13b,Nakamura15,Yu15,Ivanov16a,Rafi16}.
Most of them focus on the region around the delta resonance, with the dynamical coupled-channels model of Ref.~\cite{Nakamura15} being an exception that, by means of the unitarization of the amplitude, is able to provide predictions at somewhat larger invariant masses ($W\lesssim2$ GeV).

Recently, we have presented a model for SPP off the nucleon that aims at providing a uniform description of the reaction over the broad energy range active in neutrino experiments~\cite{Gonzalez-Jimenez17}. 
This model cures some pathologies present in many of the microscopic models commonly used, which exhibit a nonphysical behavior in the high-energy regime.
The starting point in Ref.~\cite{Gonzalez-Jimenez17} was a low-energy model that contains the $s$- and $u$-channel diagrams of the $P_{33}(1232)$ (delta), $D_{13}(1520)$, $S_{11}(1535)$, and $P_{11}(1440)$ resonances and the tree-level background terms derived from chiral perturbation theory (ChPT) for the $\pi N$ system~\cite{Hernandez07,Hernandez13,Scherer12}. 
The high-energy behavior was obtained in a Regge-based approach, where the $t$-channel Feynman propagators from the background terms were replaced by the corresponding Regge trajectories~\cite{Guidal97,Kaskulov10a}. 
Finally, the low- and high-energy models were combined in a phenomenological way into a hybrid model that can be used in the entire energy region. 

This hybrid model~\cite{Gonzalez-Jimenez17}, developed for SPP off the nucleon, is extended here to the case of incoherent SPP on the nucleus. 
We use the impulse approximation to simplify the treatment of the hadronic current, i.e., we assume that the neutrino couples to a single nucleon in the nucleus. We describe the bound nucleon wave functions using the Relativistic Mean-Field model (RMF)~\cite{Walecka74,Serot86,Serot92,Ring96}.
The RMF model provides a microscopic description of the ground state of the nucleus that is consistent with quantum mechanics, special relativity and the symmetries of the strong interaction. 
It starts from a Lorentz-covariant Lagrangian containing the nucleon and the $\sigma$- and $\omega$-meson fields. The interaction is described by the exchange of point-like mesons between point-like nucleons. Then, approximating the fields by their mean values, a mean field is generated. 
Finally, the wave function of the bound nucleon is obtained in the Hartree approximation, i.e., it is a solution of the Dirac equation in the presence of self-consistent vector (repulsive) and scalar (attractive) strong potentials with spherical symmetry. 
The parameters (coupling constants of the mesons and the mass of the $\sigma$ meson) that describe the nucleon-nucleon interaction are fit to reproduce general properties of nuclear matter and of some finite well-known spherical nuclei, such as the mean charge radius, binding energy, and neutron density profile.

As mentioned, in this work the bound nucleons are represented by RMF wave functions.
The outgoing nucleon and pion, however, are described by plane waves, i.e., final-state interactions (FSI) are ignored. 
This approach is usually referred to as the Relativistic Plane-Wave Impulse Approximation (RPWIA). 
Although this implies an important simplification of the problem, the results provided by the RPWIA serve as fundamental tests in the development of more sophisticated models. 

In Refs.~\cite{Gonzalez-Jimenez14b,Megias16a}, it is shown that the RPWIA describes the QE peak well for inclusive $(e,e')$ processes when the momentum transfer $q$ (in the laboratory frame) is larger than $600$-$700$ MeV. 
This typically corresponds to kinematic conditions in which the momentum of the outgoing nucleon is large; therefore, the effect of the distortion due to the interaction with the residual nucleus is expected to be small. 
For slow nucleons, however, the distortion of the nucleon wave function significantly modifies the cross section and should not be neglected. These effects appear in exclusive $(e,e'p)$ as well as inclusive $(e,e')$ cross section calculations~\cite{Udias93,Udias95,Udias01,Maieron03,Caballero05,Butkevich07,Meucci09,Gonzalez-Jimenez13c,Martini16}.
In the past, the RPWIA has been employed to study effects associated with the off-shell character of nucleons in nuclei, gauge ambiguities in the current operator, the role played by the lower components in the nucleon wave functions and the use of relativistic versus non-relativistic operators. 
For instance, in Refs.~\cite{Caballero98a,Caballero98b}, the RPWIA was used to study the QE $(e,e'p)$ reaction, in Ref.~\cite{Gonzalez-Jimenez15c} the helicity asymmetry in $(\vec{e},e'p)$ was analyzed, and in Ref.~\cite{Fernandez-Ramirez08} the RPWIA was applied to pion photoproduction on oxygen.

The RPWIA has previously been applied to charged-current neutrino-induced incoherent SPP in Ref.~\cite{Praet09}. 
This model, which included only the delta pole, was extended in Ref.~\cite{Gonzalez-Jimenez16a} to incorporate the $D_{13}(1520)$ resonance and the background contributions from ChPT. 
Working on that base, here we implement the more complete SPP model of Ref.~\cite{Gonzalez-Jimenez17} in the RMF framework. 
Some advantages of our approach, as compared to others, are summarized below:
\begin{itemize}
 \item The process is described in a fully relativistic framework. Both kinematic and dynamic relativistic effects, related to the structure of the operators and the lower components of the nucleon wave functions, are naturally implemented.
 \item As a consequence of the model used, in-medium effects like Fermi motion and binding energies are consistently included.
 \item We work at the amplitude level, and therefore, we provide predictions for both inclusive and exclusive processes. 
 \item The elementary pion-production vertex is described with the hybrid model of Ref.~\cite{Gonzalez-Jimenez17}, where the high-energy behavior of the amplitude is given by a Regge approach. This allows us to provide predictions in the entire $W$ region, from the pion threshold to high invariant masses ($W$ larger than 2 GeV). This contrasts with conventional low-energy models that show a nonphysical behavior in the high-energy regime.
\end{itemize}
The RPWIA is the first building block for relativistic models aiming at predicting electron and neutrino scattering processes. 
Still, the elastic and inelastic FSI are missing.~\footnote{By elastic FSI we refer to those mechanisms in which particles are not created or absorbed in the nucleus. The rest are inelastic FSI.}
We are working to implement the elastic distortion of the outgoing hadrons within a relativistic and consistent quantum mechanical approach (Sec.~\ref{sec:hadronic}).  

Modeling of the inelastic part of the FSI is a very challenging task. The fact that it is not possible to control the kinematics within neutrino experiments, as traditionally done in electron scattering, greatly complicates the interpretation of the experimental cross sections. 
Inelastic FSI such as charge-exchange reactions, pion absorption, pion production in secondary interactions, etc., along with the contributions from deep-inelastic scattering (DIS), may affect the multiplicity of visible hadrons in the final state. 
To the best of our knowledge, currently the only way of approaching this problem is using MC generators, that generally employ cascade models for describing the FSI and PYTHIA routines~\cite{Pythia6} for the hadronization in DIS.
Some generators widely used in the literature are NuWro~\cite{Golan12}, GENIE~\cite{Andreopoulos10}, NEUT~\cite{Hayato09}, and GiBUU~\cite{Mosel15b}. The latter, based on quantum-kinetic transport theory, is an exception regarding the treatment of FSI.

Very often, however, the description of the elementary vertices in these generators is oversimplified or treated in such a pragmatic way that it is difficult to disentangle what is the actual level of understanding of the physical processes. 
In this sense, the predictions of microscopic models (such as the one presented here) may serve as a test of our current knowledge of the fundamental interaction.

Finally, a consistent description of the world data set of neutrino-nucleus pion-production cross sections is still missing, and some open questions remain to be answered. 
For example, the predictions from GiBUU~\cite{Lalakulich13b} and Hernandez et al.~\cite{Hernandez13}, that fairly match with each other for the $\nu$-induced $\pi^+$ production at MiniBooNE kinematics~\cite{MBCCpionC11}, do not agree with the MiniBooNE data~\cite{MBCCpionC11} when FSI are included. On the contrary, these data are reproduced well when FSI are ignored. 
Since the outgoing pion is no doubt still interacting with the residual nucleus, this is obviously a problem that requires further investigation. 
Another related issue is the apparent inconsistency between the MiniBooNE and MINERvA data; this topic was further discuss in Refs.~\cite{Sobczyk15,Mosel17}.

The structure of this paper is as follows. 
In Sec. II, the kinematics and cross section formula are presented. In Sec. III, we describe the hadronic current within our approach. We compare our results with MiniBooNE and MINERvA data, as well as with NuWro predictions in Sec. IV. In Sec. V, we present our conclusions.

\section{Kinematics and Cross section}\label{sec:kin-xs}

\begin{figure}[htbp]
  \centering  
      \includegraphics[width=0.35\textwidth,angle=0]{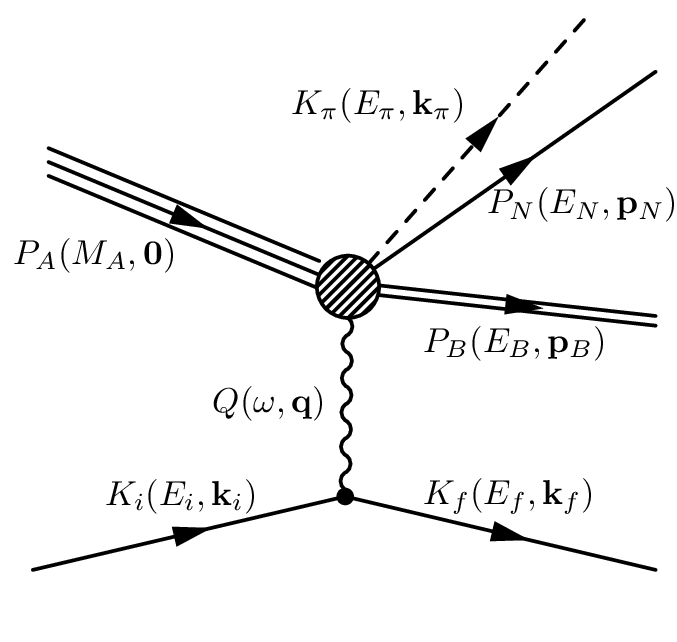}
      \vspace{-0.5cm}
  \caption{Feynman diagram describing the electroweak (incoherent) SPP process.
  An incoming lepton with 4-momentum $K_i$ interacts with a nucleus at rest, $P_A$, by the exchange of a single boson $Q$. This results in a scattered lepton $K_f$, the residual nucleus $P_{A-1}$, and an outgoing nucleon $P_N$ and pion $K_\pi$.
}
  \label{kinematic1}
\end{figure}

We focus on the modeling of the process shown in Fig.~\ref{kinematic1}.   
The exclusive cross section describing this process is
\fl{
 \frac{\de^{8}\sigma}{\de E_f\de\Omega_f \de E_\pi \de\Omega_\pi \de\Omega_N} = 
	{\cal F}\ \frac{k_f E_f p_N^2E_\pi k_\pi}{(2\pi)^{8}f_{rec}}\ l_{\mu\nu}h^{\mu\nu}.\label{XS}
}
This expression applies for both electron (electromagnetic interaction) and neutrino (weak-neutral current and charge-current interactions) induced SPP. 
The function $f_{rec}$ accounts for the recoil of the residual nucleus and is given by
\ba
  f_{rec} = \frac{p_N}{E_N}\left(1 + \frac{E_N}{E_{A-1}}\left|1 +\frac{\np_N\cdot(\nk_\pi-\nq)}{p_N^2}\right| \right)\,.
\ea
The leptonic tensor $l_{\mu\nu}$ and the factor ${\cal F}$, which includes the boson propagator as well as the coupling constants of the leptonic vertex, were defined in Ref.~\cite{Gonzalez-Jimenez17}. 
The hadronic tensor $h^{\mu\nu}$ is described in detail in the next section. 

In what follows, we provide some details on the kinematics of the process.
The scattering process is completely determined by 9 independent variables. 
We chose the laboratory variables: $ E_i$, $ E_f$, $\theta_f$, $\phi_f$, $E_\pi$, $\theta_\pi$, $\phi_\pi$, $\theta_N$, $\phi_N$. 
The $\hat{\nz}$ axis is defined here along the direction of the incident beam $\nk_i$ ($\hat{\nz}\parallel\nk_i$).
This choice of reference frame is different from the usual procedure in which $\hat{\nz}\parallel\nq$.
Working in the usual reference frame ($\hat{\nz}\parallel\nq$) allows one to decompose the cross section in terms of hadronic response functions, which simplifies the analysis of the problem.
In particular, in Ref.~\cite{Donnelly85}, it was shown that in the two-particle knockout case, if one applies the change of variables $(\phi_\pi,\phi_N)\mapsto(\phi,\Delta\phi)$ with $\phi=\phi_\pi+\phi_N$ and $\Delta\phi=(\phi_\pi-\phi_N)/2$, the dependence of the nuclear responses on $\phi$ factorizes in terms of sine and cosine functions. 
The advantage of this is that the integral over $\phi$ can be done analytically.
In this work, however, we want to study differential cross sections as functions of the pion scattering angle relative to the direction of the incident beam. In that case, the integral over $\phi$ cannot be performed analytically.

We consider the residual nucleus as a bound system that can be in an excited state.  
Its mass, $M_{A-1}$, is determined from the relation 
\fl{
  E_m &= M_{A-1} + M -M_A\,, \label{Em}
}
where $M$ is the free nucleon mass and $E_m$ is the missing energy. For $E_m$ we use empirical values that depend on the shell in which the hole was created.

The previous considerations, along with energy-momentum conservation, allow us to determine all 4-vectors involved in the scattering process.

\section{Hadronic Tensor}\label{sec:hadronic}

The hadronic tensor is defined as
\ba
  h^{\mu\nu} = \frac{1}{2j+1}\sum_{m_j,s_N} (J^\mu)^\dagger J^\nu\,,
\ea
where $j$ is the total angular momentum of the bound nucleon, its third component is $m_j$, and $s_N$ is the spin projection of the outgoing nucleon.
$J^\mu$ represents the expected value of the hadronic current. Within the relativistic impulse approximation, it has the general structure 
\ba
  J^\mu \sim \psib_N\ \phib\ {\cal O}_{1\pi}^\mu\ \psi\,.  \label{HT}
\ea
$\psi$ and $\psib_N$ ($\equiv\psi_N^\dagger\gamma_0$) are Dirac spinors describing the bound and scattered nucleons, and $\phi$ is the wave function of the pion. 
${\cal O}_{1\pi}^\mu$ represents the hadronic current operator that induces the transition between the initial one-nucleon state and the final one-nucleon one-pion state. 
In this work, we use the same current operator ${\cal O}_{1\pi}^\mu$ that was developed in Ref.~\cite{Gonzalez-Jimenez17} for the description of the electroweak SPP on the free nucleon.
The explicit expressions and more details about the model can be found in Ref.~\cite{Gonzalez-Jimenez17}. 

\begin{figure}[htbp]
  \centering  
      \includegraphics[width=.4\textwidth,angle=0]{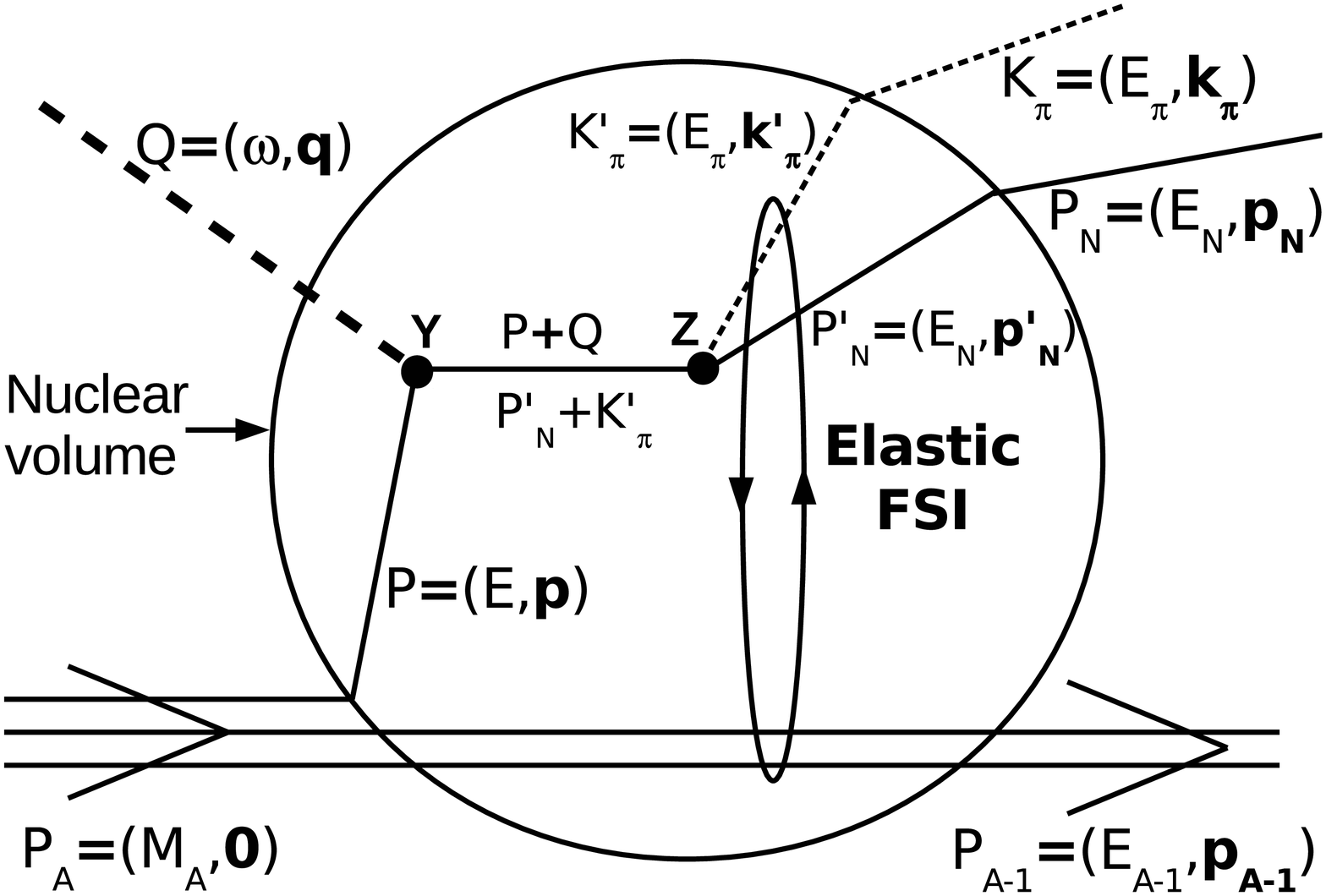}
      \vspace{-0.3cm}
  \caption{Representation of an $s$-channel-like diagram within the impulse approximation. See text for details. }
  \label{NP-FSI}
\end{figure}

We discuss now how to consistently describe the hadronic current when the reaction occurs inside the nucleus. For that, in Fig.~\ref{NP-FSI} we show the case of an $s$-channel like diagram, taken as an example.
In the interaction vertex $Y$, a virtual boson $Q$ couples to a bound nucleon $P$ which propagates as a nucleon or resonance with 4-momentum $Q+P$. 
In the interaction vertex $Z$, the virtual baryon `decays' into a nucleon and a pion with $P_N'$ and $K_\pi'$, respectively. 
Inside the nuclear volume, energy-momentum conservation reads $Q+P = P_N' + K_\pi'$.

We consider the particles as energy eigenstates; therefore, their energies are the same outside and inside the nuclear volume.
The momentum of the particles inside the nuclear volume, however, is given by a probability distribution (nuclear wave functions), i.e., none of the particles are on-shell.
In the case of the outgoing hadrons, this can be expressed by explicitly including the dependence of the wave functions on the asymptotic and local momenta, i.e., $\psi=\psi(\np_N',\np_N)$ and $\phi=\phi(\nk_\pi',\nk_\pi)$.
This treatment of the wave functions of the outgoing hadrons, which includes distortion effects due to the presence of the residual nucleus, is the only way to account for the elastic FSI in a consistent, fully relativistic and quantum-mechanical way. 
Still, one would need to account for the inelastic FSI.  

In this framework, the hadronic current reads
\fl{
 &J^\mu = \int{\de\np_N'} \int{\frac{\de\np}{(2\pi)^{3/2}}}\times\non\\
  & \psib_{s_N}(\np_N',\np_N) \phib(\nk'_\pi,\nk_\pi) {\cal O}_{1\pi}^\mu(Q,K'_\pi,P'_N) \psi_\kappa^{m_j}(\np), \label{J2-general}}
with $K'_\pi = Q+P-P_N'$. Note that ${\cal O}_{1\pi}^\mu(Q,K'_\pi,P'_N)$ depends on the local variables. The Dirac spinor of the bound nucleon, $\psi_\kappa^{m_j}$, is labeled by the quantum numbers $m_j$ and $\kappa$, the latter being related to the total angular momentum by $j=|\kappa|-1/2$. 
 
The implementation in this model of the elastic distortion of the pion and the nucleon wave functions is under development. In the present work, however, we concentrate solely on the aforementioned RPWIA approach.

\subsection{Hadronic current within the RPWIA}\label{sec:had-current}

Within the RPWIA, the elastic distortion of the outgoing nucleon and the pion is ignored, i.e., they are described as plane waves. 
In that case, $K'_\pi=K_\pi$ and $P'_N=P_N$, and the hadronic current can be written as
\ba
  J^\mu &=& {\cal N}\, \int\de\nr\, e^{i(\nq-\np_N-\nk_\pi)\cdot\nr}\non\\
      &\times&\ubarN {\cal O}_{1\pi}^\mu(Q,K_\pi,P_N) \psi_\kappa^{m_j}(\nr)\,,\label{J1-RPWIA}
\ea
where $u$ represents a free Dirac spinor and ${\cal N}=\sqrt{M/(2E_\pi E_N)}$ is the normalization factor for the outgoing nucleon and pion plane waves.
Since in Eq.~\ref{J1-RPWIA} the only dependence on $\nr$ appears in the bound wave function, this expression can be further reduced by introducing the Fourier transform of the bound-nucleon wave function:
\fl{
   J^\mu = (2\pi)^{\frac{3}{2}} {\cal N}\, \ubarN  {\cal O}_{1\pi}^\mu(Q,K_\pi,P_N)\, \psi_\kappa^{m_j}(\np),\label{J-Fourier} 
}
with $\np=\np_N+\nk_\pi-\nq$. 

We use the Oset and Salcedo parametrization to account for the modification of the delta-decay width due to in-medium effects~\cite{Hernandez13,Oset87,Nieves93,Gil97}. We will refer to this as OSMM (Oset and Salcedo medium modifications). 
The delta-decay width in the OSMM formula is a function of the center-of-mass pion kinetic energy and the nuclear density $\rho(r)$. 
Therefore, the hadronic current operator becomes an $r$-dependent function and the full 3-dimensional integral over $\nr$ in Eq.~\ref{J1-RPWIA} cannot be performed analytically. Still, using the properties of spherical harmonics, one can analytically resolve the angular dependence $\de\Omega_{\nr}$.
The impact of the medium modification of the delta width on the cross section is investigated and described in more detail in the next section.

\section{Results}

In this section, 
we present a systematic comparison of our model predictions with the MiniBooNE and MINERvA charged-current SPP data.
In the MiniBooNE $\nu$-induced $1\pi^+$ ($1\pi^0$) production sample~\cite{MBCCpionC11,MiniBooNECCpi010}, 
the experimental signal is defined as any event with a $\pi^+$ ($\pi^0$) and a $\mu^-$ detected in coincidence, with no other visible mesons. The average neutrino energy is around 1 GeV, 
in the $\pi^+$ sample it ranges from 0 to 3 GeV while the $\pi^0$ sample is restricted to the region 0.5-2 GeV.
In the MINERvA $\nu$-induced $1\pi^+$ production sample~\cite{MINERvACCpi15}, 
the signal definition is less restrictive. In this case, events with exactly one muon and one charged pion exiting the nucleus are accepted, with no limitation on neutral pions or other mesons.
A cut on the `experimental invariant mass', $W_{exp}<1.4$ GeV
~\footnote{The experimental invariant mass is defined as~\cite{MINERvACCpi15,MINERvACCpi16} 
\ba
W_{exp}=\sqrt{M^2+2M^2(E_\nu-E_\mu)-Q^2}\,,\label{Wexp}
\ea
with  $Q^2=E_\nu( E_\mu-k_\mu\cos\theta_\mu)-m_\mu^2$. 
Note that $W_{exp}$ coincidences with the `true invariant mass' $W=\sqrt{(P+Q)^2}$, only when the target nucleon is on-shell and at rest.
To implement the cuts, we use the variable $W_{exp}$.}, 
is applied to focus on the delta region.
The flux goes from 1.5 to 10 GeV, with an average energy of $\sim$4 GeV. 
In the MINERvA $\bar\nu$- and $\nu$-induced $1\pi^0$ production samples~\cite{MINERvACCpi16,MINERvAnuCC17}, the signal is defined as only one $\pi^0$ exiting the nucleus, with no other mesons detected. 
The antineutrino flux extends from 1.5 to 10 GeV, with average energy of $\sim$4 GeV, 
while the neutrino flux is the same as in the 1$\pi^+$ sample but including the high-energy tail up to 20 GeV. 
The cut $W_{exp}<1.8$ GeV is applied in both neutrino and antineutrino samples.
Our calculations include the possibility of scattering off hydrogen, which is present in the MiniBooNE (CH$_2$) and MINERvA (CH) targets.

We want to stress that RPWIA predictions do not account for any elastic or inelastic FSI mechanisms, beyond those included in the OSMM of the delta width. 
Therefore, we do not aim at reproducing the MiniBooNE and MINERvA data. 
Instead of that, our goal is to provide an accurate microscopic description of the elementary reaction, which has to be the core of any reliable prediction of the experimental data.

\subsection{Medium modification of the delta-decay width}\label{sec:MM}

The decay width of the delta resonance is modified inside the nucleus. 
As is widely done in the literature~\cite{Hernandez13,Sobczyk13,Lalakulich13b,Martini14,Gonzalez-Jimenez16a}, we use the Oset and Salcedo prescription to evaluate these medium modifications. 

The procedure consists in replacing the free-delta width by its in-medium value (see Ref.~\cite{Hernandez13} and references therein for details), i.e.,
\ba
\Gamma^{free} &\longrightarrow& \Gamma^{med} = \Gamma_{PB} - 2\Im(\Sigma_\Delta)\,.\label{GammaMM}
\ea
The Pauli blocking term in Eq.~\ref{GammaMM}, $\Gamma_{PB}$, accounts for the fact that some of the nucleons from the delta decay may be Pauli blocked, decreasing the decay width.
$\Im(\Sigma_\Delta)$ is the imaginary part of the delta self-energy, which, in the OSMM formula, contains contributions from three different processes: 
i) $\Delta N \rightarrow \pi N N$, 
ii) $\Delta N \rightarrow N N$,
and iii) $\Delta N N \rightarrow N N N$.
This means that new decay channels beyond the $\Delta\rightarrow\pi N$ are now opened, increasing the delta width.
The net effect, resulting from the competition of Pauli blocking and the three new decay channels, is an (energy-dependent) increase of the width.~\footnote{The delta mass receives contributions from the real part of the delta self-energy. 
We follow the same approach as in~\cite{Hernandez13} and ignore these corrections. 
Its impact on the cross sections is relatively small compared with other uncertainties.}

\begin{figure}[htbp]
  \centering  
      \includegraphics[width=.25\textwidth,angle=270]{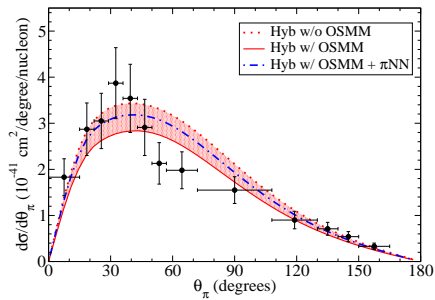}
  \caption{MINERvA $\nu$-induced $1\pi^+$ production sample~\cite{MINERvACCpi15} compared with RPWIA predictions. The solid (dotted) line is the result with (without) medium modification of the delta width. The dash-dotted line is the result with OSMM when the contribution from the $\Delta N \rightarrow \pi N N$ channel is added to the cross section. The results were computed with the hybrid model (see Sec.\ref{sec:lem-hyb}).  }
  \label{fig:OSMM} 
\end{figure}

The decay channel $\Delta N \rightarrow \pi N N$ contributes to the MiniBooNE and MINERvA pion-production signal, therefore, when it is included in the delta width, it should also be added incoherently to the cross section. 
In Ref.~\cite{Hernandez13}, this process was roughly modeled as the delta-pole amplitude multiplied by a weighting factor. 
This adds extra strength to the cross section and brings the results closer to the ones without medium modification.
Obviously, this is far from being a satisfactory description of such a process and, since its contribution to the cross section is significant (Fig.~\ref{fig:OSMM}), one should be cautious when interpreting the results.

The validity of the OSMM formula is limited to a range given approximately by $T_\pi^*<300$ MeV, with $T_\pi^*$ being the pion kinematic energy in the delta rest frame. 
The kinematics involved in the neutrino-nucleus reactions that are investigated here span a broader kinematic region, bringing more uncertainties to the reliability of the procedure.

Another issue is the lack of consistency of using the OSMM prescription in our model. 
The OSMM formula was developed in a particular framework: a Fermi-gas-based model. 
In infinite nuclear matter (Fermi gas), the nucleons are labeled only by their momentum, as a consequence, only the struck nucleons with momentum above the Fermi momentum can be knocked out. Thus, the Pauli blocking is necessary and has to be added ad hoc.  
In a mean-field framework (such as the one used here) the bound nucleons belong to shells labeled with different quantum numbers. 
Therefore, as long as the energy transferred to a bound nucleon is larger than its binding energy, the nucleon will be knocked out independently of its momentum.
In summary, Pauli blocking should not be implemented in a shell model, at least not in the same way as in a Fermi-gas model.

For the reasons explained above, we consider the medium modification of the delta width as an uncertainty in our model.
We present our results computed with and without medium modification; they can presumably be interpreted as an upper and lower limit, respectively.
In Fig.~\ref{fig:OSMM}, we illustrate this by plotting a red band. The dash-dotted line that lies within the band is the result with OSMM when the $\Delta N \rightarrow \pi N N$ contribution is added as described in Ref.~\cite{Hernandez13}.

\subsection{Low-energy model vs hybrid model}\label{sec:lem-hyb}

\begin{figure*}[htbp]
  \centering  
      \includegraphics[width=.23\textwidth,angle=270]{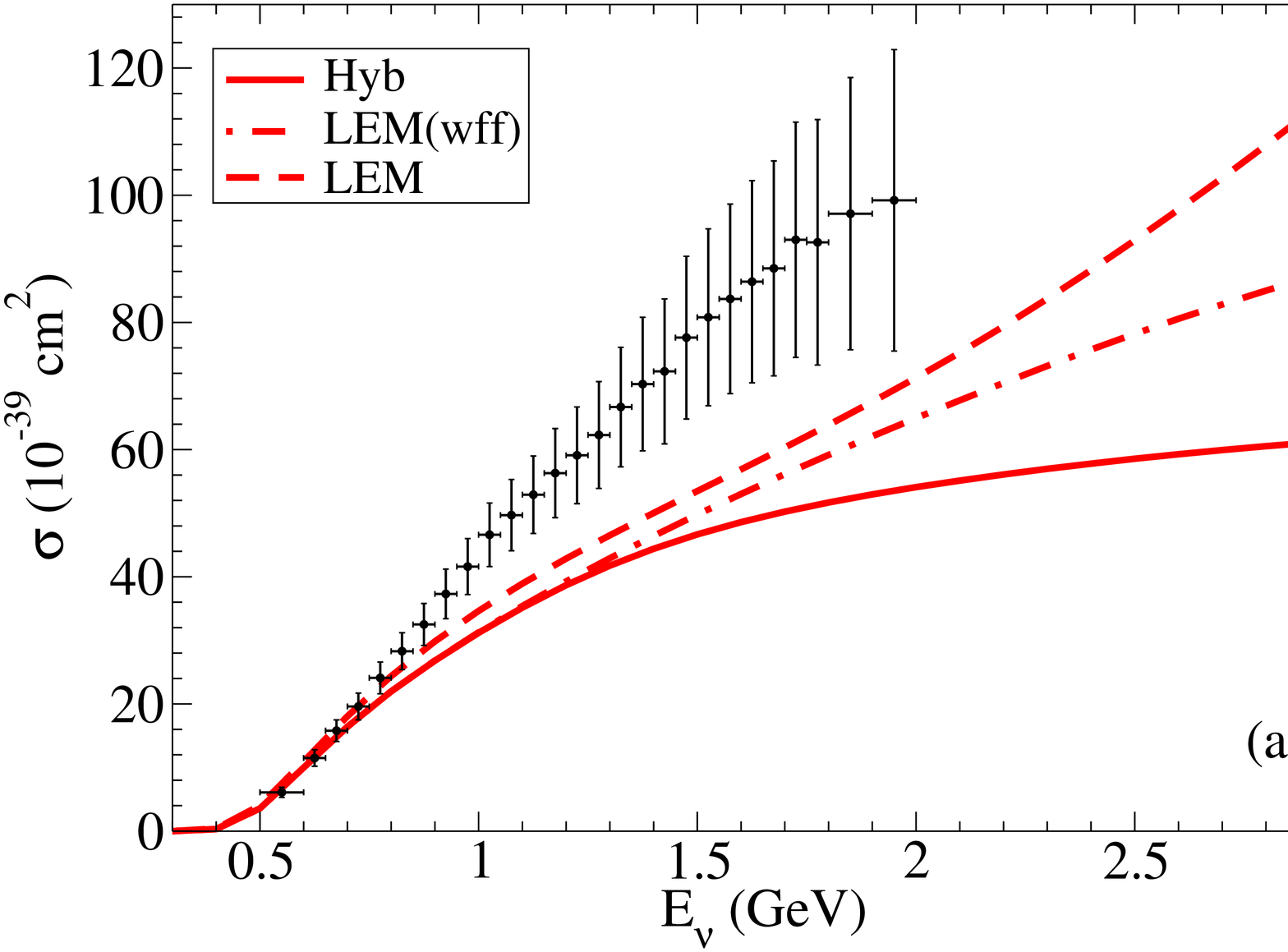}
      \includegraphics[width=.23\textwidth,angle=270]{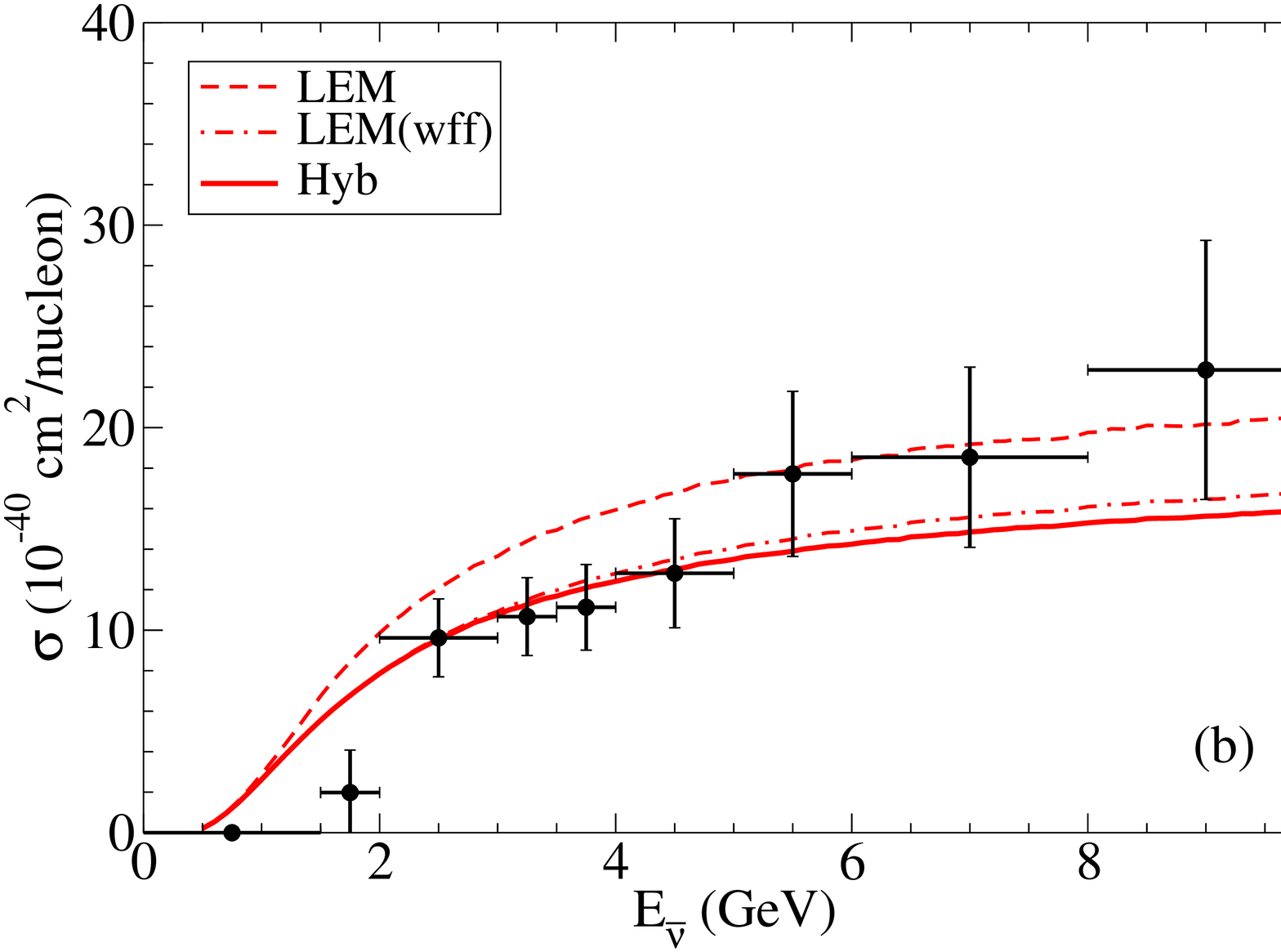}
      \includegraphics[width=.24\textwidth,angle=270]{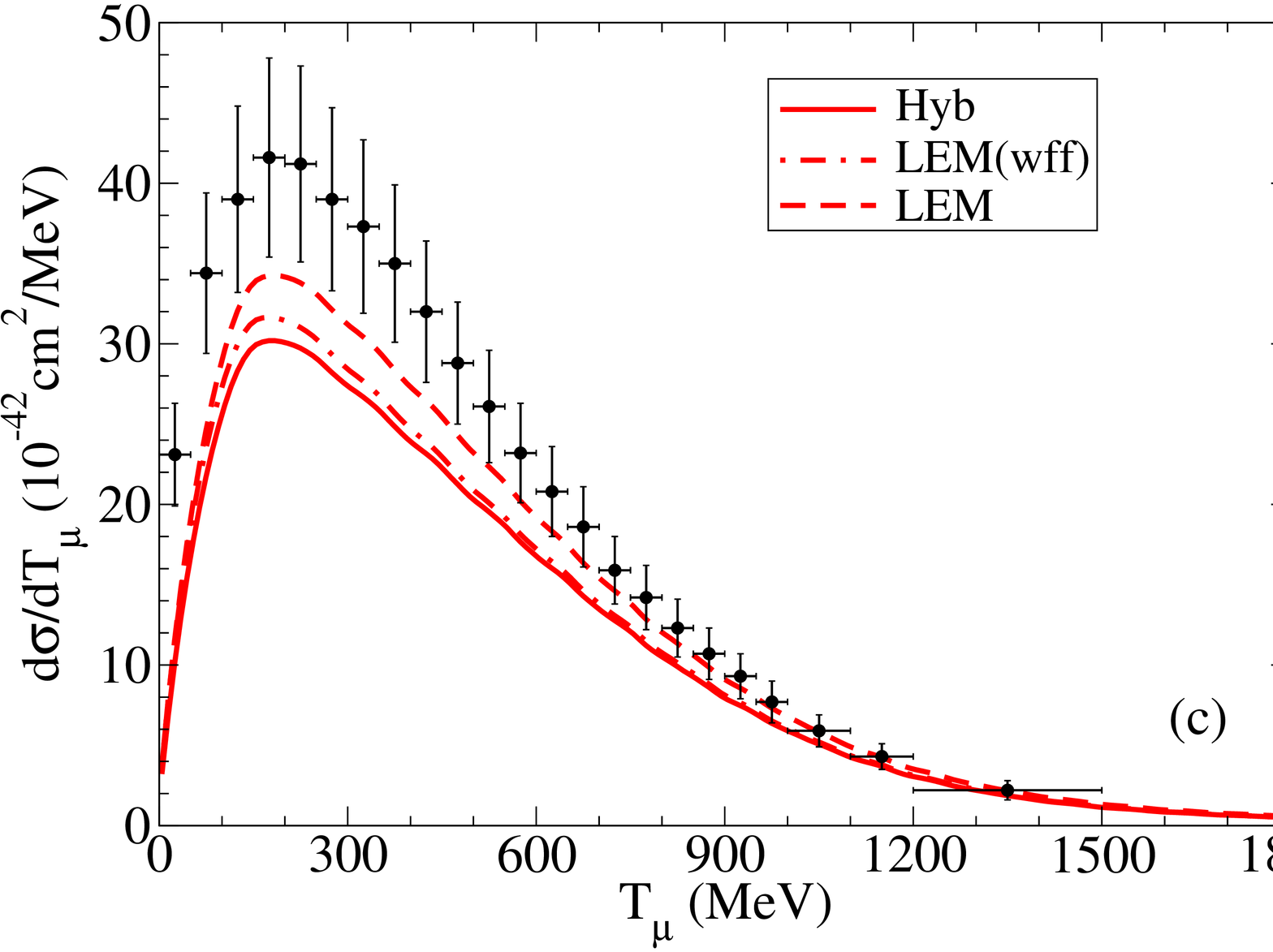}
      \includegraphics[width=.24\textwidth,angle=270]{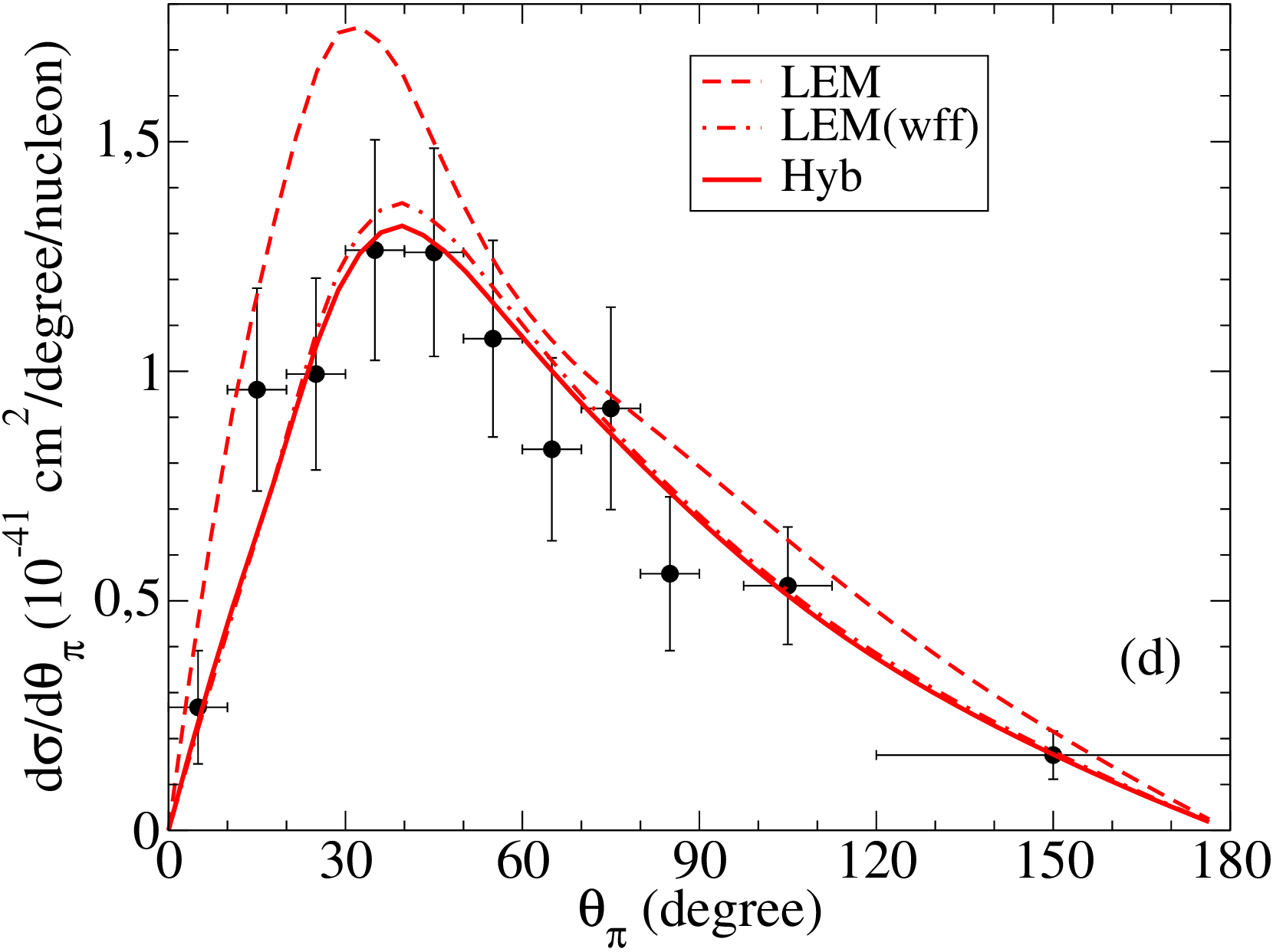}
  \caption{In the left panels we compare our predictions with the MiniBooNE neutrino $1\pi^+$~\cite{MBCCpionC11} data. The right panels correspond to the MINERvA antineutrino $1\pi^0$~\cite{MINERvACCpi16} case. In panels (a) and (b), we show the total cross section. In panels (c) and (d), we present the flux-folded differential cross sections as a function of the muon kinetic energy and pion scattering angle, respectively. 
  All the results include OSMM.}
  \label{fig:hyb-lem} 
\end{figure*}

In this section, we study the impact that nonphysical strength in the amplitude, coming from the high-$W$ region, may have on the cross sections for the MiniBooNE and MINERvA kinematics.
For that, in Fig.~\ref{fig:hyb-lem} we show the predictions obtained within the three approaches summarized below (see Ref.~\cite{Gonzalez-Jimenez17} for more details):
\begin{itemize}
 \item Low-energy model (LEM, dashed-red lines):  It contains direct- and cross-channel amplitudes for the resonances $P_{33}(1232)$, $D_{13}(1520)$, $S_{11}(1535)$ and $P_{11}(1440)$, and the background terms from the ChPT $\pi N$ Lagrangian.
 \item Low-energy model with cutoff form factors [LEM(wff), dash-dotted red lines]: This is the same as LEM but including phenomenological cutoff form factors in the $s$- and $u$-channel amplitudes of the resonances. 
 This is done to regularize the pathological behavior of the amplitudes in the kinematic regions far from the resonance peak, $W\approx M_R$, where $M_R$ is the mass of the resonance. 
 \item Hybrid model (Hyb, solid-red lines): At low invariant masses ($W<1.4$ GeV), it provides exactly the same response as LEM(wff). 
 For $W>2$ GeV, it reproduces the behavior given by our Regge approach. 
 In the transition region, $1.4<W<2$ GeV, the amplitude results from a compromise between LEM(wff) and the Regge-based predictions.
\end{itemize}

In Figs.~\ref{fig:hyb-lem}(a) and \ref{fig:hyb-lem}(b), we compare our predictions with total cross sections from MiniBooNE neutrino $1\pi^+$ and MINERvA antineutrino $1\pi^0$, respectively.
In Fig.~\ref{fig:hyb-lem}(a), the nonphysical behavior of the background terms starts to show up at $E_\nu>1.2$ GeV, and at $E_\nu=2$ GeV the difference between the LEM(wff) and the hybrid model reaches $\sim17\%$. This is due to the fact that no cut on the invariant mass is applied; therefore, large $W$ values can contribute to the cross section for large neutrino energies.
In Fig.~\ref{fig:hyb-lem}(b), the cutoff form factors in the resonances notably reduces the cross section, while the nonphysical strength from the background terms is small due to the cut $W_{exp}<1.8$ GeV.

The MiniBooNE flux peaks at relatively low energies; therefore, the differences between the three models are reduced when studying the flux-folded differential cross sections. This is shown in Fig.~\ref{fig:hyb-lem}(c), where we present the single-differential cross section as a function of the muon kinetic energy. 
In Fig.~\ref{fig:hyb-lem}(d), we present the differential cross section folded with the MINERvA flux as a function of the pion scattering angle. 
Similar to Fig.~\ref{fig:hyb-lem}(b), the cutoff form factors in the resonances reduce notably the cross section; however, due to the cut $W_{exp}<1.8$ GeV, the nonphysical strength from the background terms is small.

\subsection{Hybrid model, NuWro and data} \label{sec:HybridvsNuWro}

The implementation of the final-state interactions between the outgoing hadrons and the residual nucleus is a fundamental ingredient for a meaningful comparison with the MiniBooNE and MINERvA pion-production data.
Indeed, due to FSI, the observed spectra will be distorted compared to the ones from the elementary reaction. 
The primary pions may be absorbed, suffer charge exchange or rescatter elastically.
Additionally, the pions that leave the nucleus may originate from other secondary interactions.
To quantitatively estimate the effect of FSI, we compare our hybrid-RPWIA results with those of the Monte Carlo neutrino event generator NuWro~\cite{NuWro-web}.

In NuWro, the elementary SPP in the region $W<1.6$ GeV is described by the delta resonance (within the Adler model~\cite{Adler75}) and an effective background extrapolated from the DIS contribution. 
For $W>1.6$ GeV, the predictions are based on the DIS formalism~\cite{Bodek02} and the PYTHIA 6 hadronization routines~\cite{Pythia6}. 
In the region $1.4 < W < 1.6$ GeV, a smooth transition between the resonance and the DIS regions is performed~\cite{Sobczyk05}.
Note that through DIS, depending on $Q^2$ and the available energy $W$, a bunch of hadrons in the final state can be created, with single-pion production being just a fraction of the total. 
Which events will contribute to the cross section will depend on the particular definition of the signal, which is different for each data sample and not free from ambiguities.
The FSI in NuWro are described within the intranuclear cascade framework~\cite{Golan12}, where the dynamics follows the Oset et al. model~\cite{Salcedo88}.
In this work, we use NuWro 17.01.1~\cite{NuWro-web} with the same recommended set of parameters introduced recently in NuWro 17.09.
Note that, because different NuWro configurations and methodology were used, the results presented here slightly differ from the ones published in Refs.~\cite{MINERvACCpi15,MINERvACCpi16,MINERvAnuCC17,Sobczyk15}.

\begin{figure*}[htbp]
  \centering  
      \includegraphics[width=.24\textwidth,angle=270]{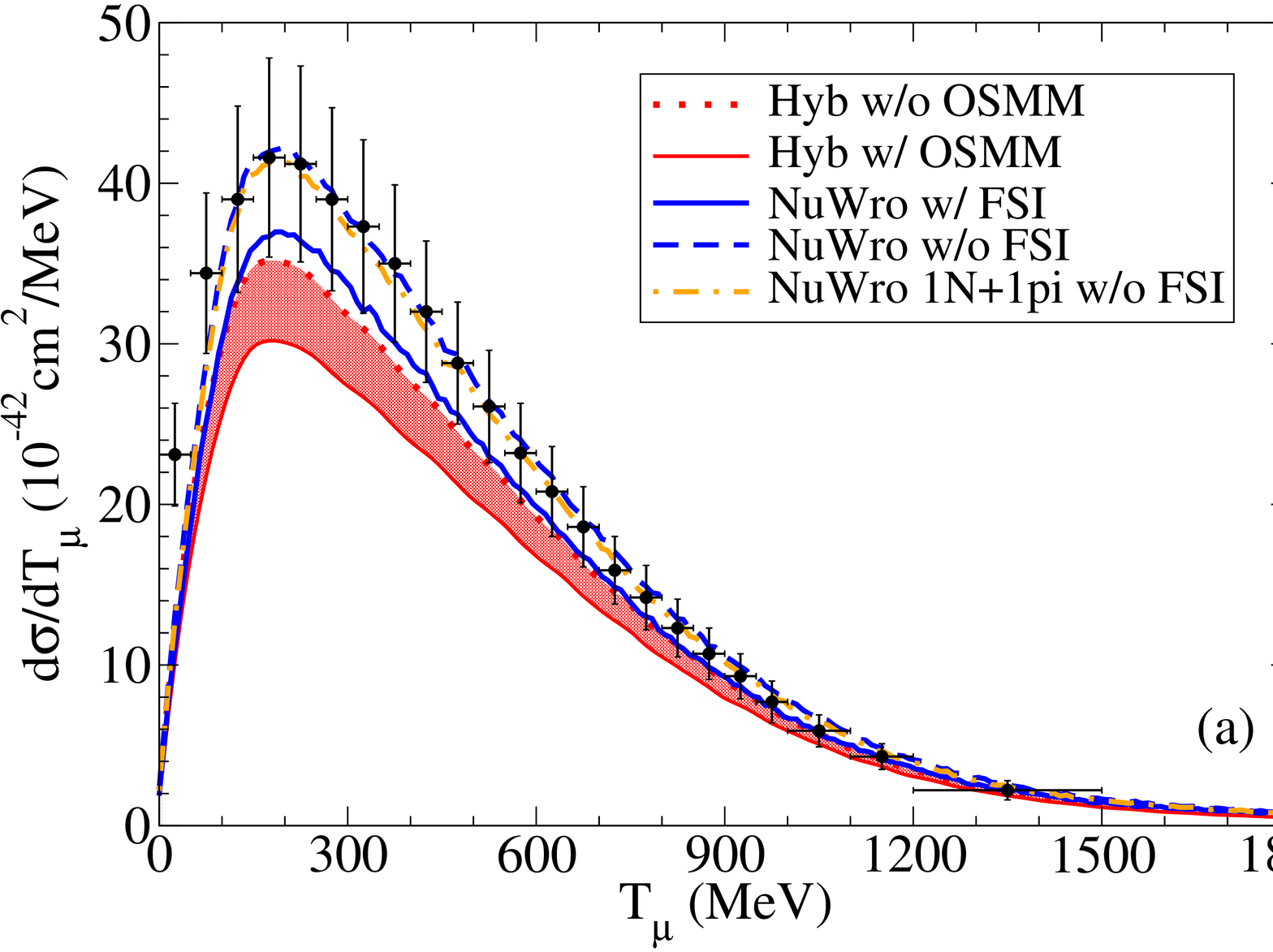}
      \includegraphics[width=.24\textwidth,angle=270]{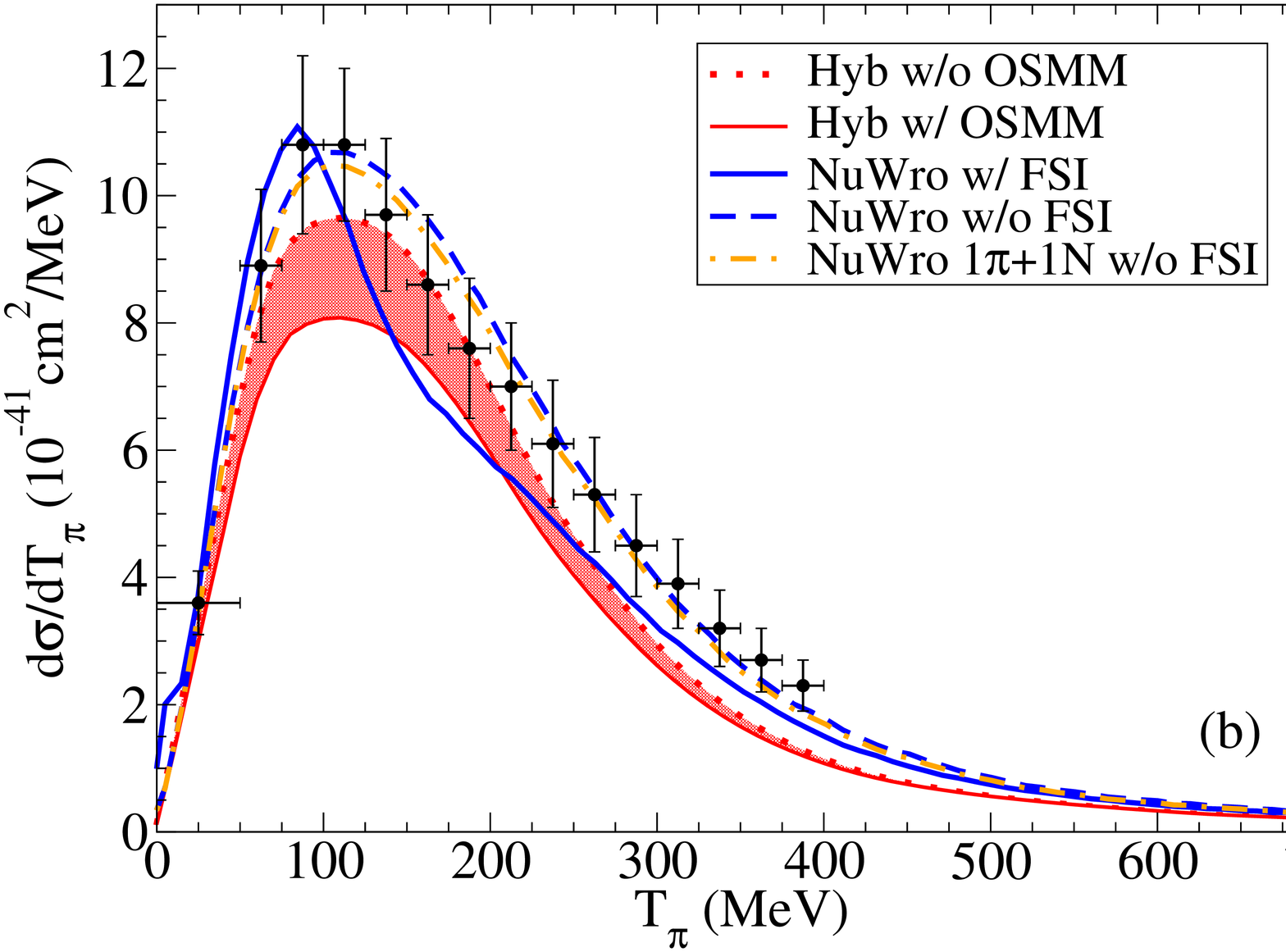}
  \caption{ Single-differential cross sections for the MiniBooNE $\nu$CC $1\pi^+$ sample~\cite{MBCCpionC11}. 
  The red bands are the hybrid-RPWIA predictions, the upper (lower) lines being the calculation with (without) medium modification of the delta width. 
  The NuWro predictions with and without FSI are shown by the blue solid and blue dashed lines, respectively. 
  The orange dash-dotted line represents the NuWro prediction of the elementary SPP process.
  The solid lines with small triangles are the hybrid-RPWIA results when the higher mass resonances P$_{11}$, S$_{11}$, and D$_{13}$ are not considered in the calculations.} 
  \label{fig:MB-pi+} 
\end{figure*}

\begin{figure*}[htbp]
  \centering  
      \includegraphics[width=.24\textwidth,angle=270]{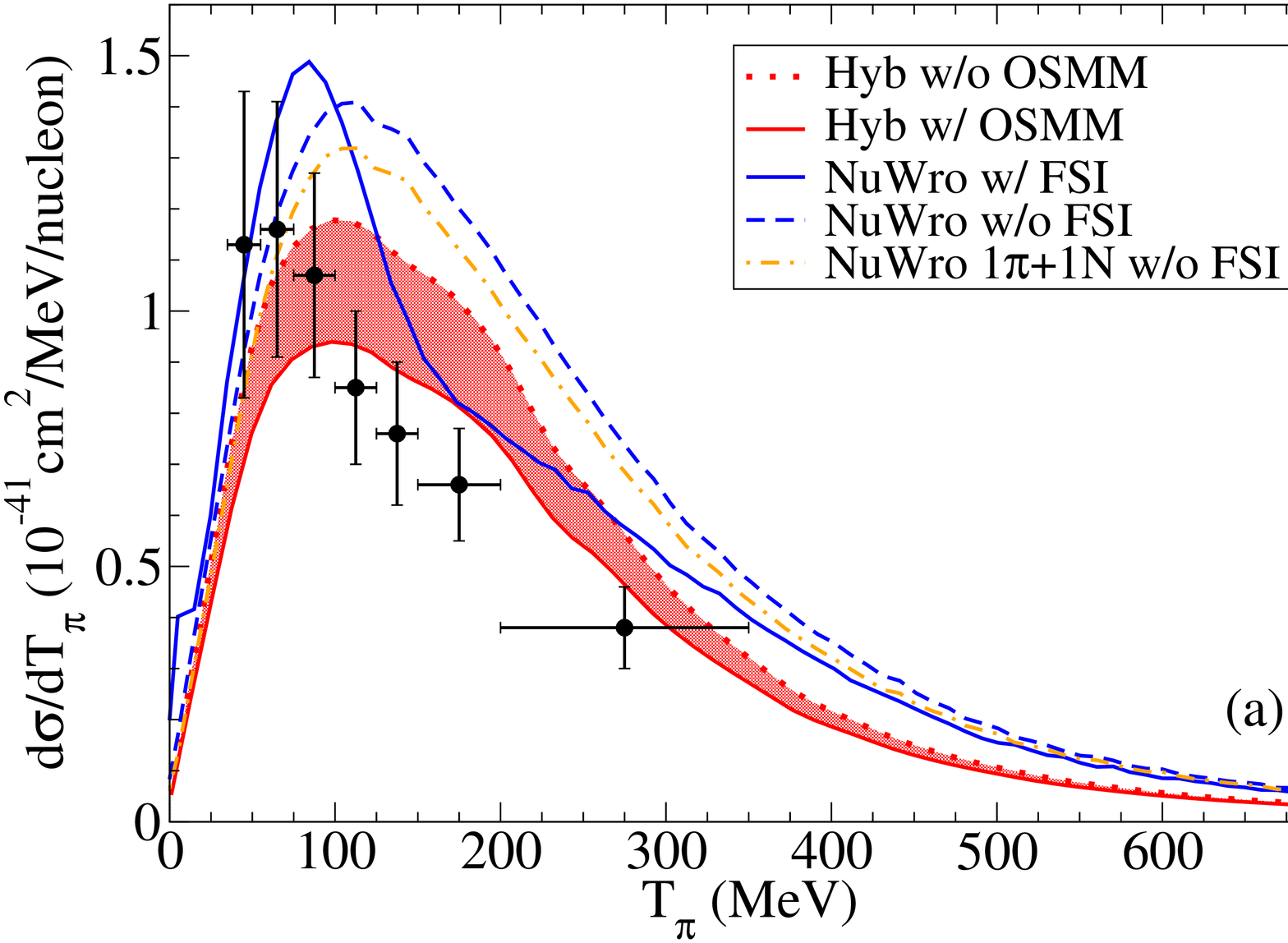}
      \includegraphics[width=.24\textwidth,angle=270]{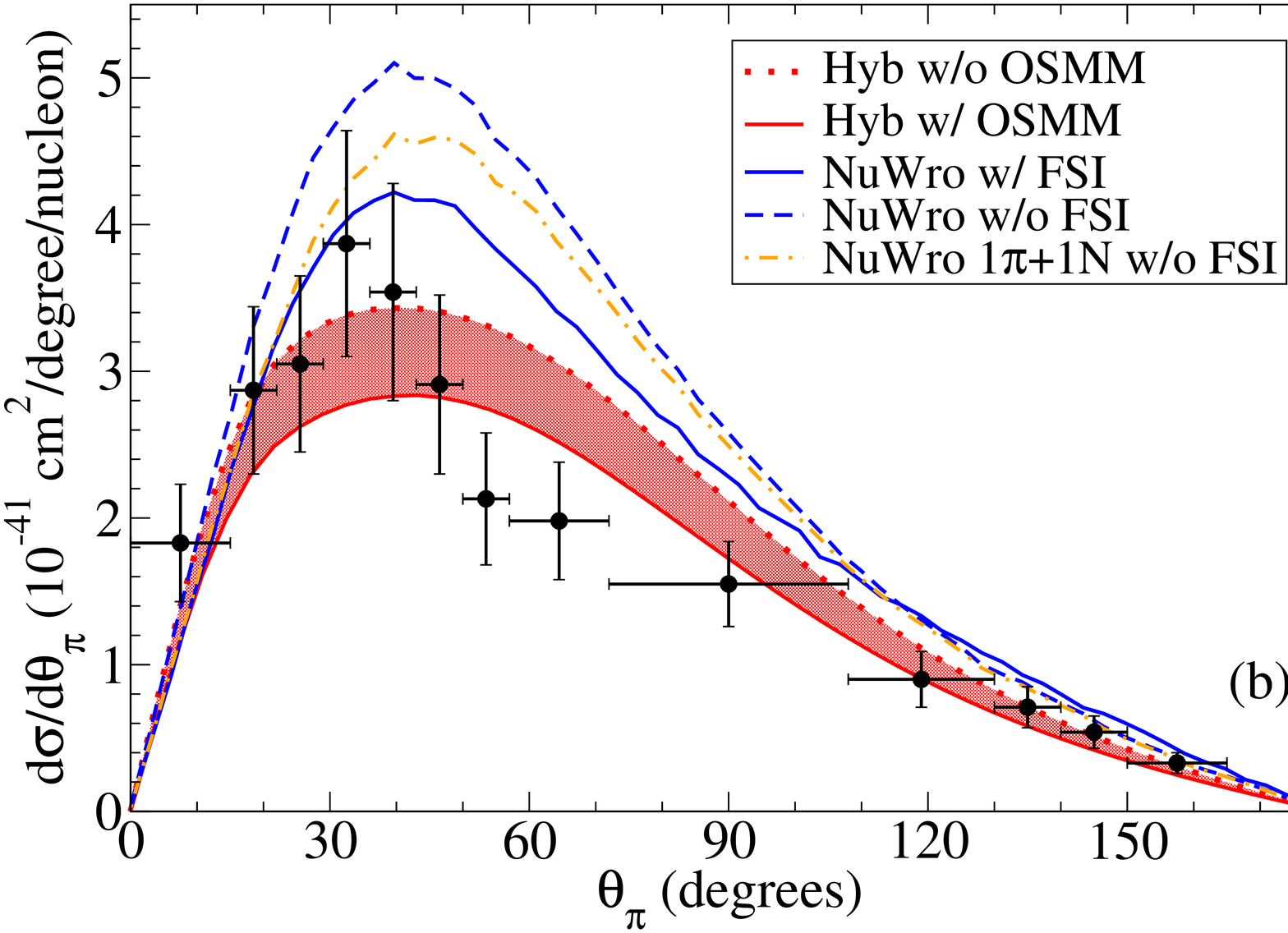}
  \caption{ Single-differential cross sections for the MINERvA $\nu$CC $1\pi^+$ sample~\cite{MINERvACCpi15}. Labels are as in Fig.~\ref{fig:MB-pi+}.}
  \label{fig:Min-pi+}   
\end{figure*}

\begin{figure*}[htbp]
  \centering  
        \includegraphics[width=.24\textwidth,angle=270]{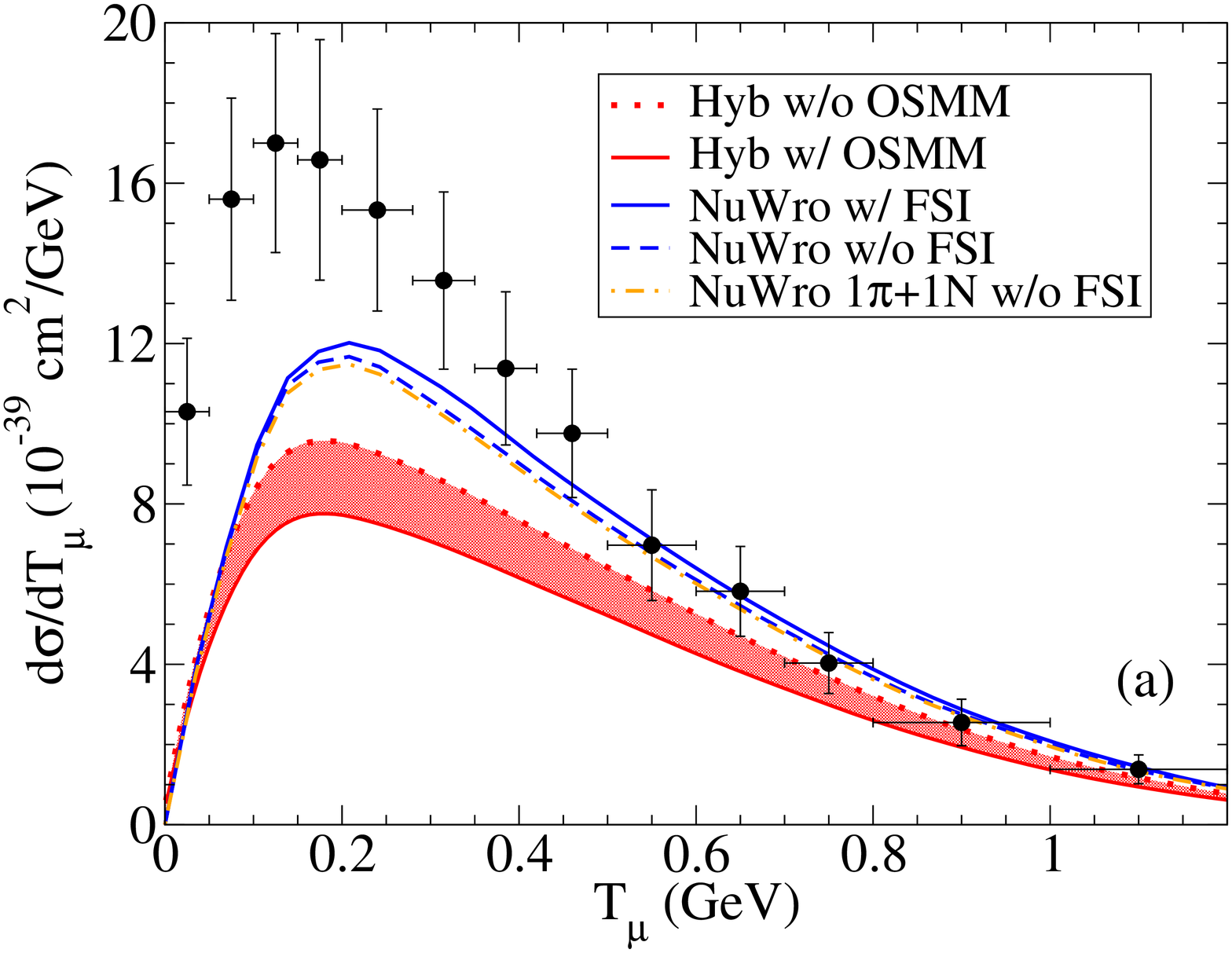}
        \includegraphics[width=.24\textwidth,angle=270]{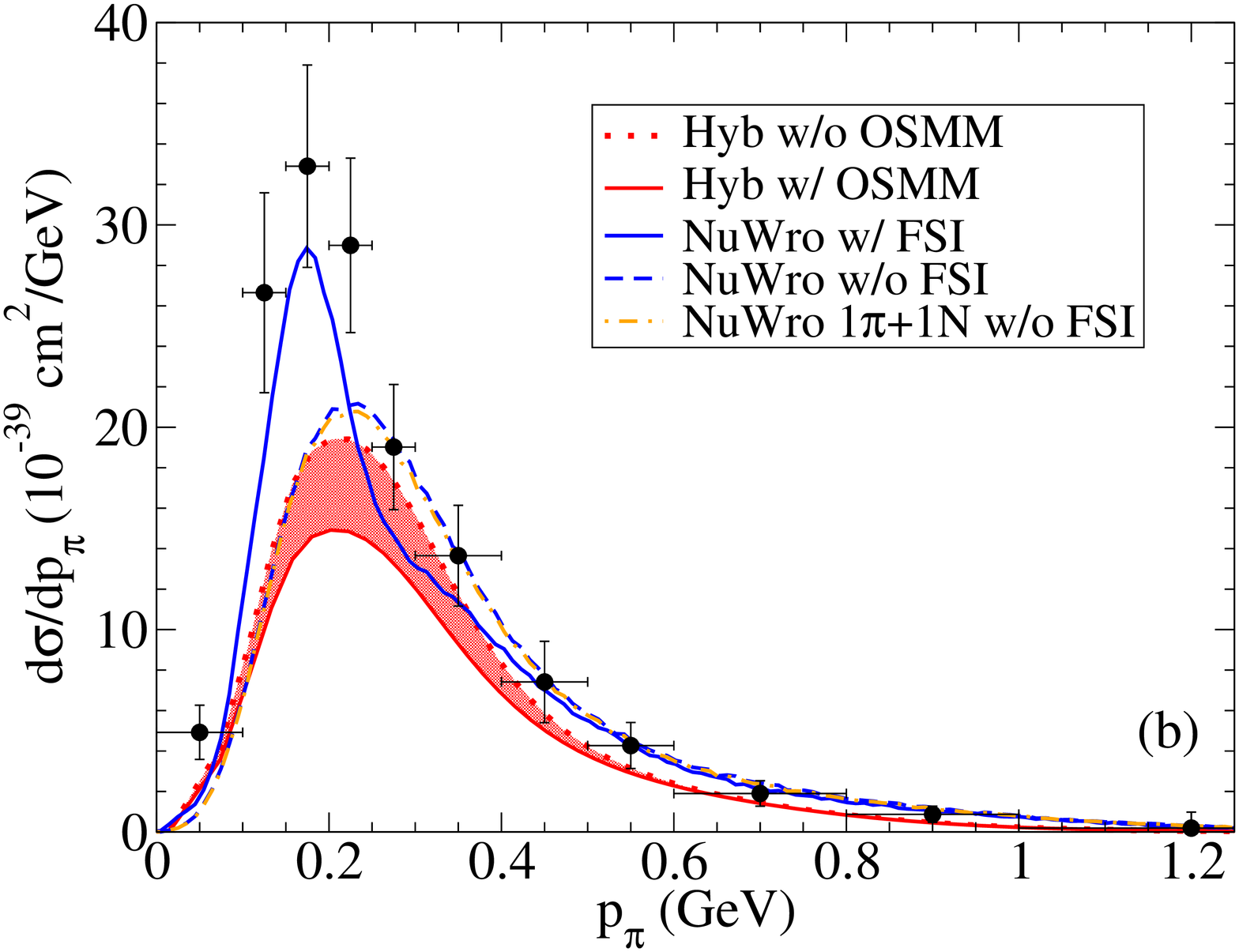}\\
        \includegraphics[width=.24\textwidth,angle=270]{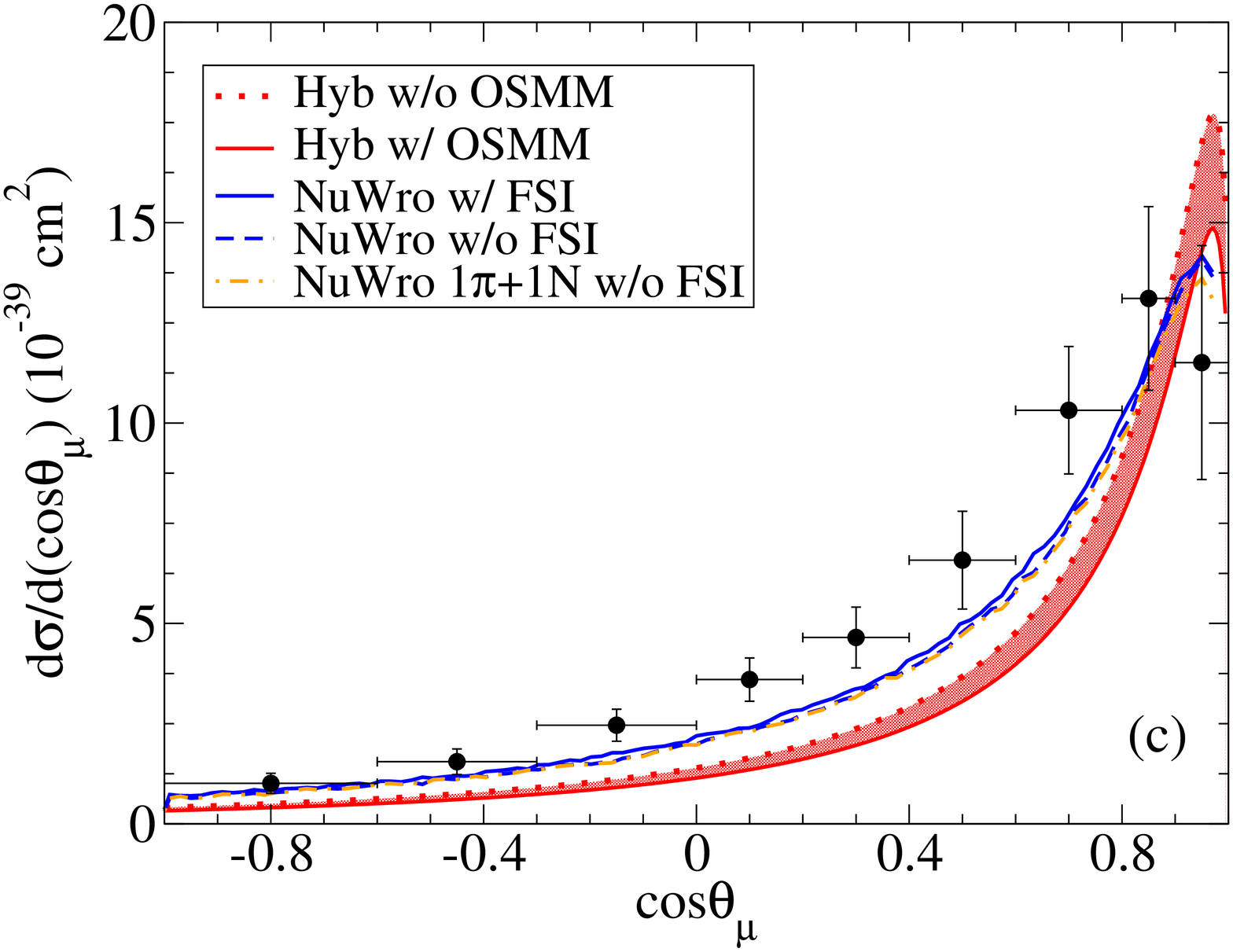}
        \includegraphics[width=.24\textwidth,angle=270]{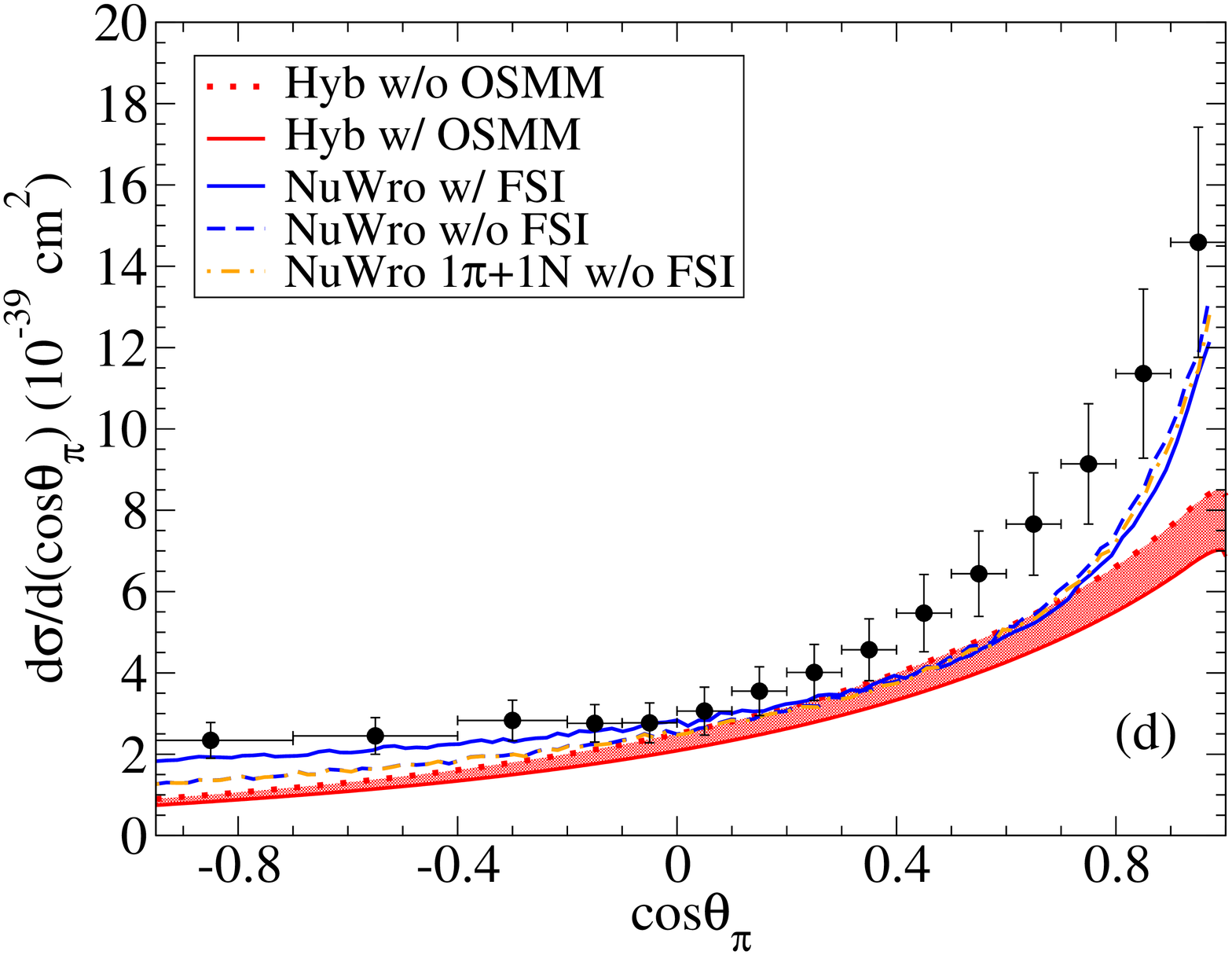}
  \caption{ Single-differential cross sections for the MiniBooNE $\nu$CC $1\pi^0$ sample~\cite{MiniBooNECCpi010}. Labels are as in Fig.~\ref{fig:MB-pi+}.}
  \label{fig:MB-pi0}   
\end{figure*}

\begin{figure*}[htbp]
  \centering  
      \includegraphics[width=.24\textwidth,angle=270]{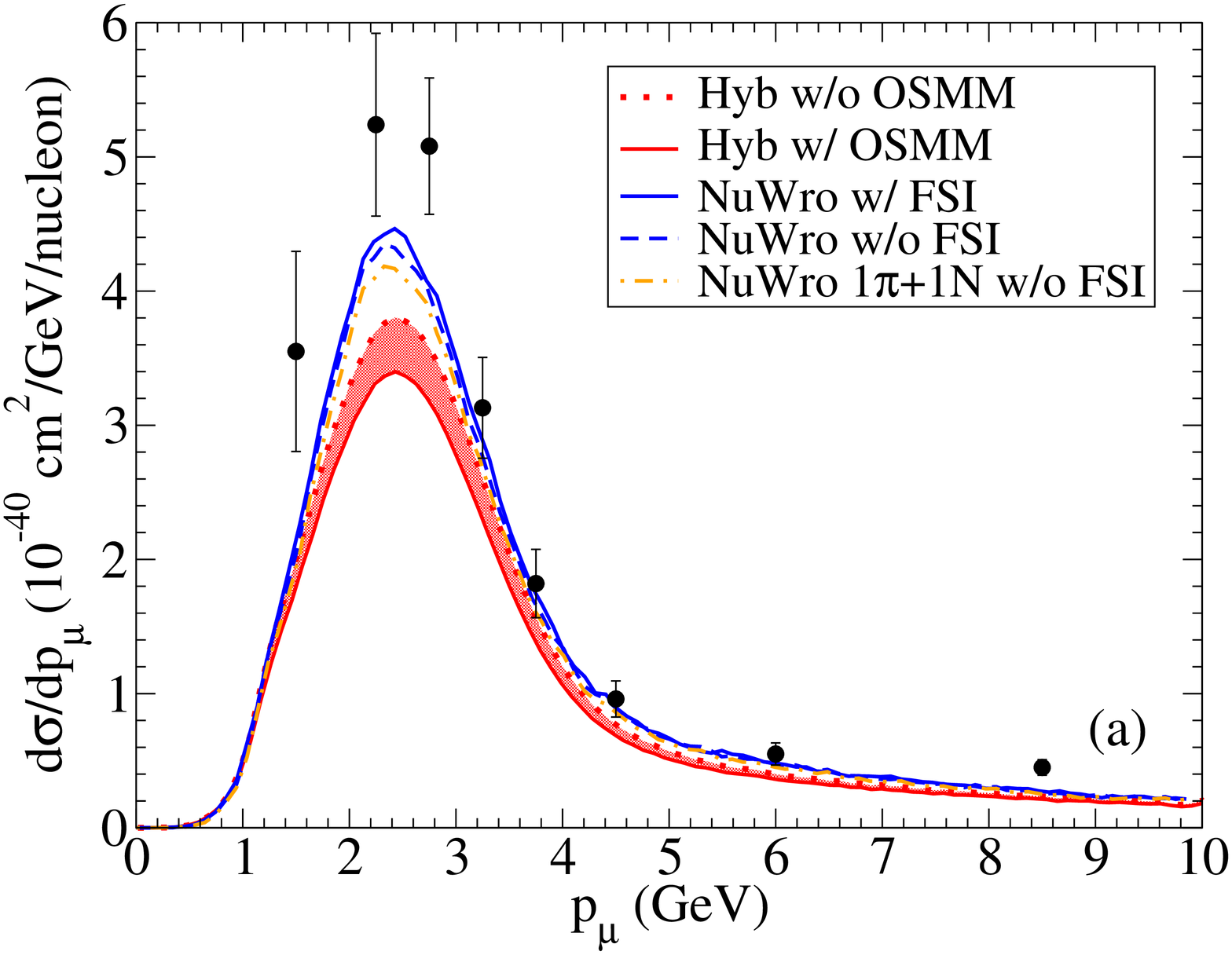}
      \includegraphics[width=.24\textwidth,angle=270]{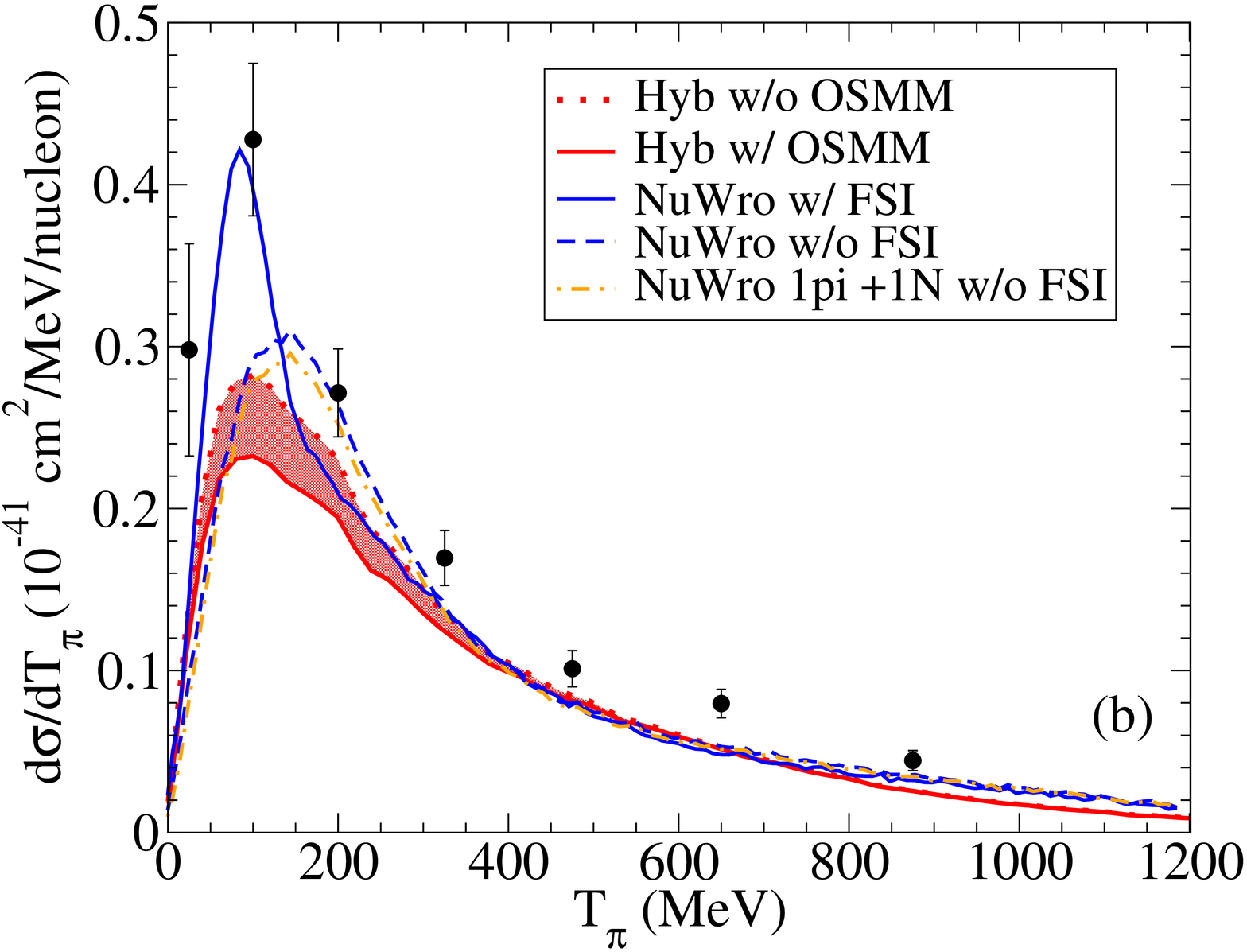}\\
      \includegraphics[width=.24\textwidth,angle=270]{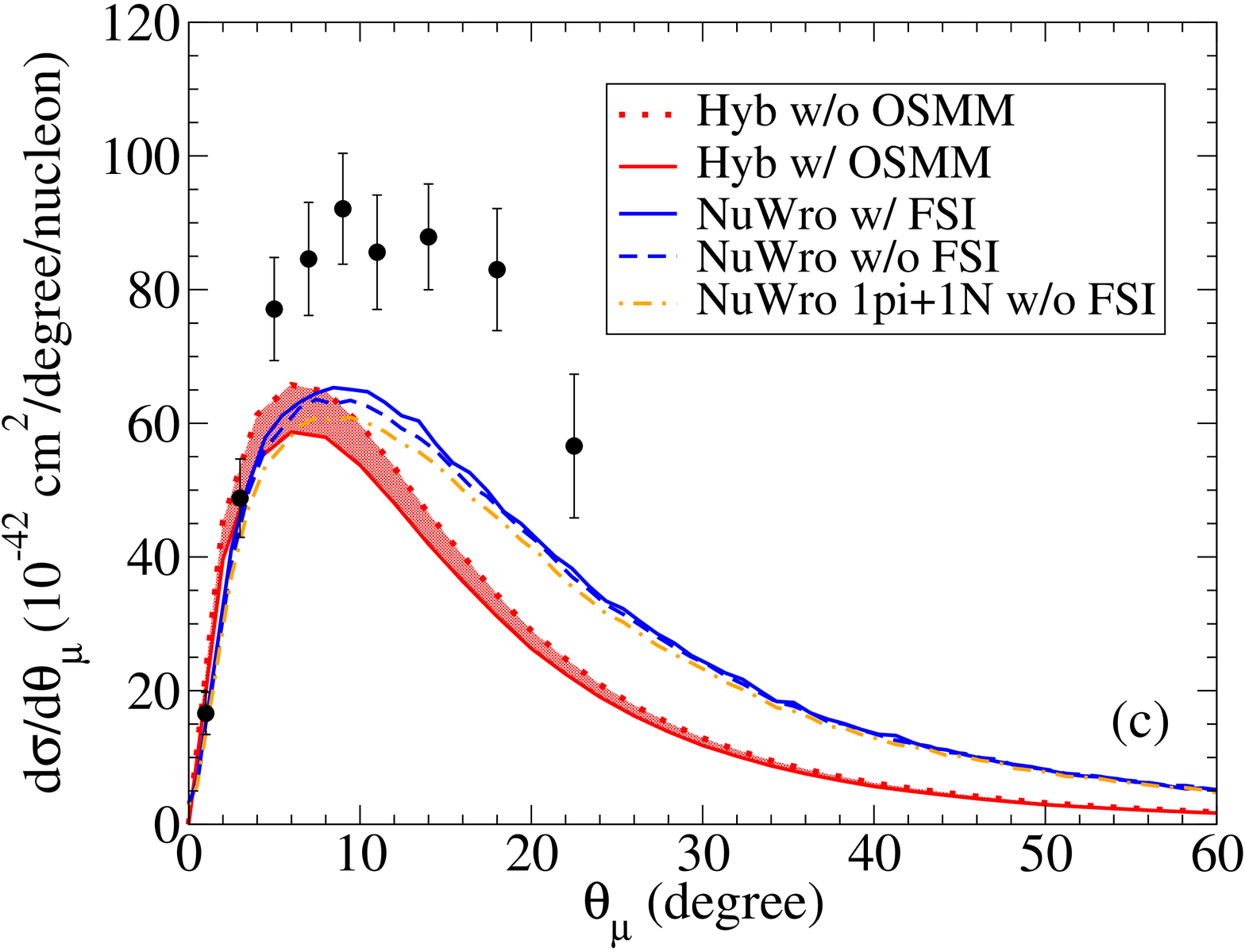}
      \includegraphics[width=.24\textwidth,angle=270]{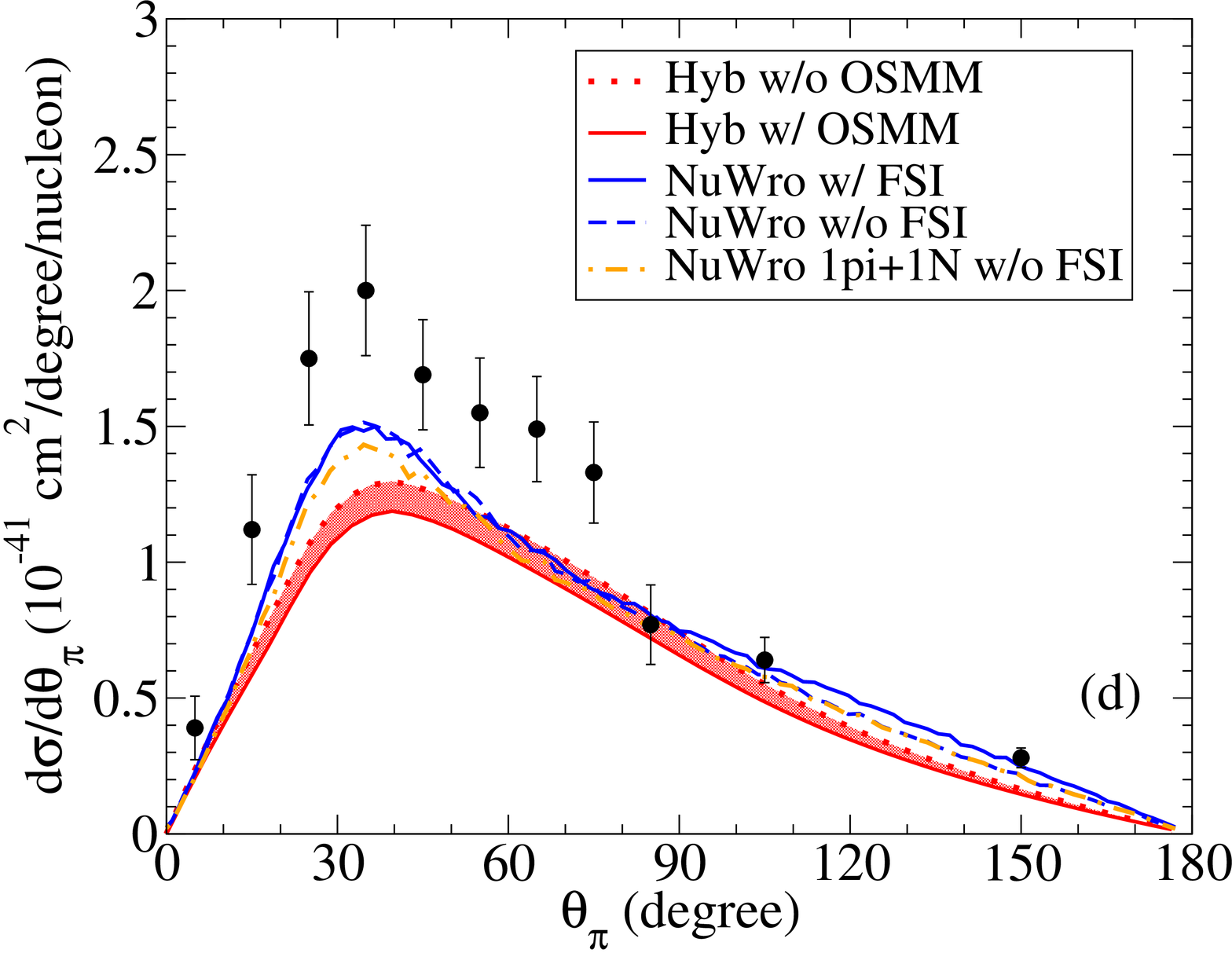}
  \caption{ Single-differential cross sections for the MINERvA $\nu$CC $1\pi^0$ sample~\cite{MINERvAnuCC17}. To mimic the experimental analysis of data~\cite{MINERvAnuCC17}, contributions only from $\theta_\mu<25$ deg are considered in the cross sections of panels (a), (b) and (d). Labels are as in Fig.~\ref{fig:MB-pi+}.}
  \label{fig:Min-nu-0pi} 
\end{figure*}

\begin{figure*}[htbp]
  \centering  
      \includegraphics[width=.24\textwidth,angle=270]{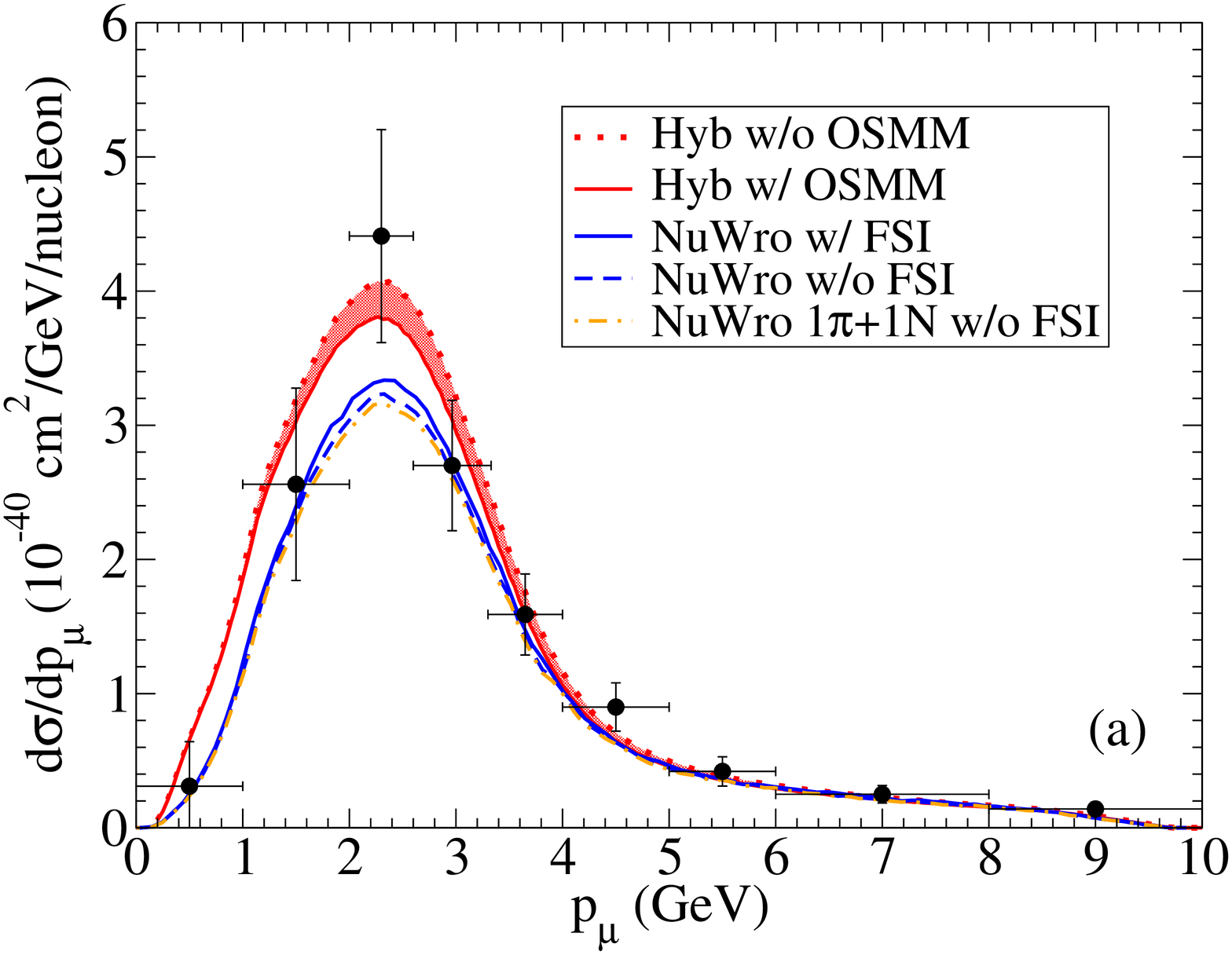}
      \includegraphics[width=.24\textwidth,angle=270]{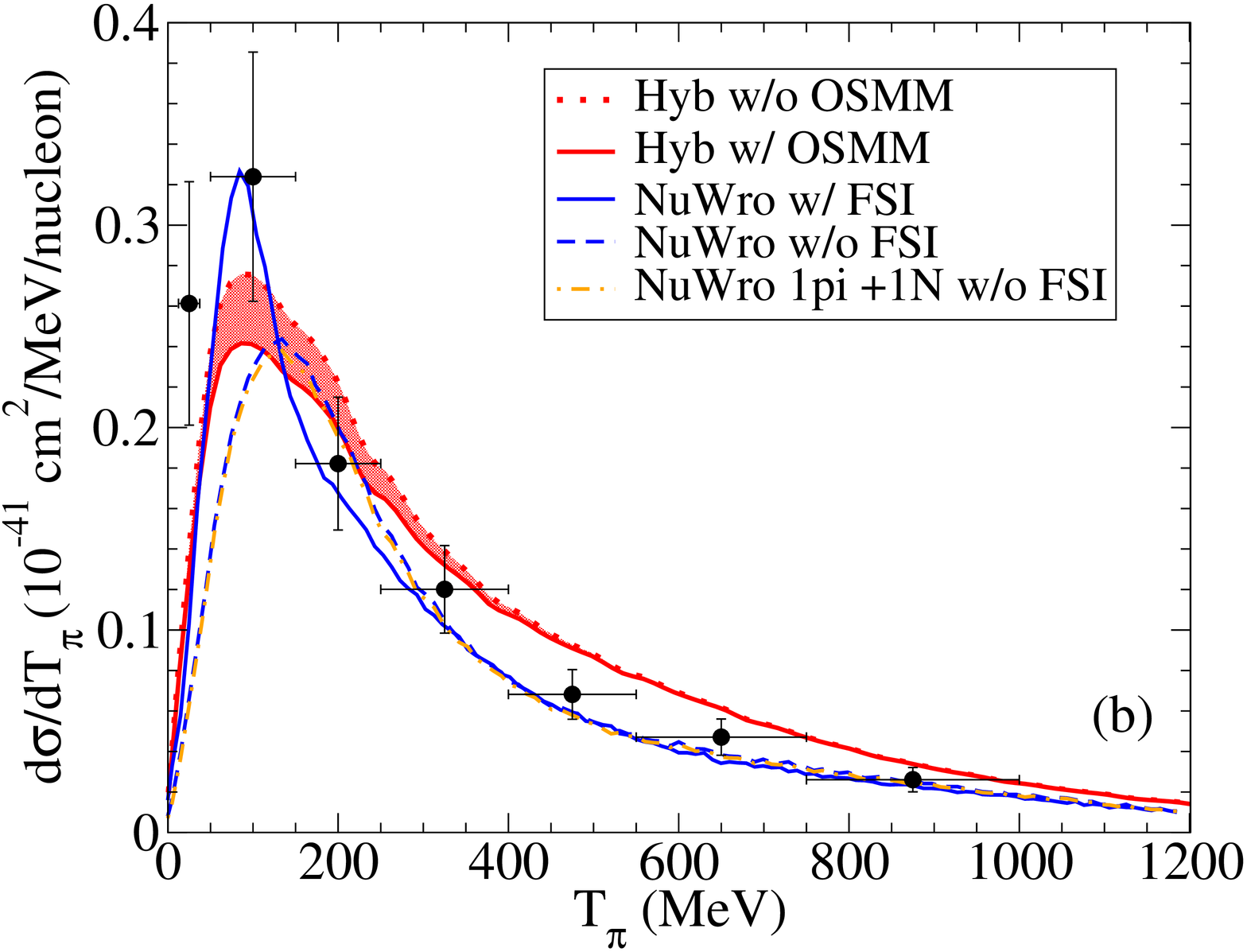}\\
      \includegraphics[width=.24\textwidth,angle=270]{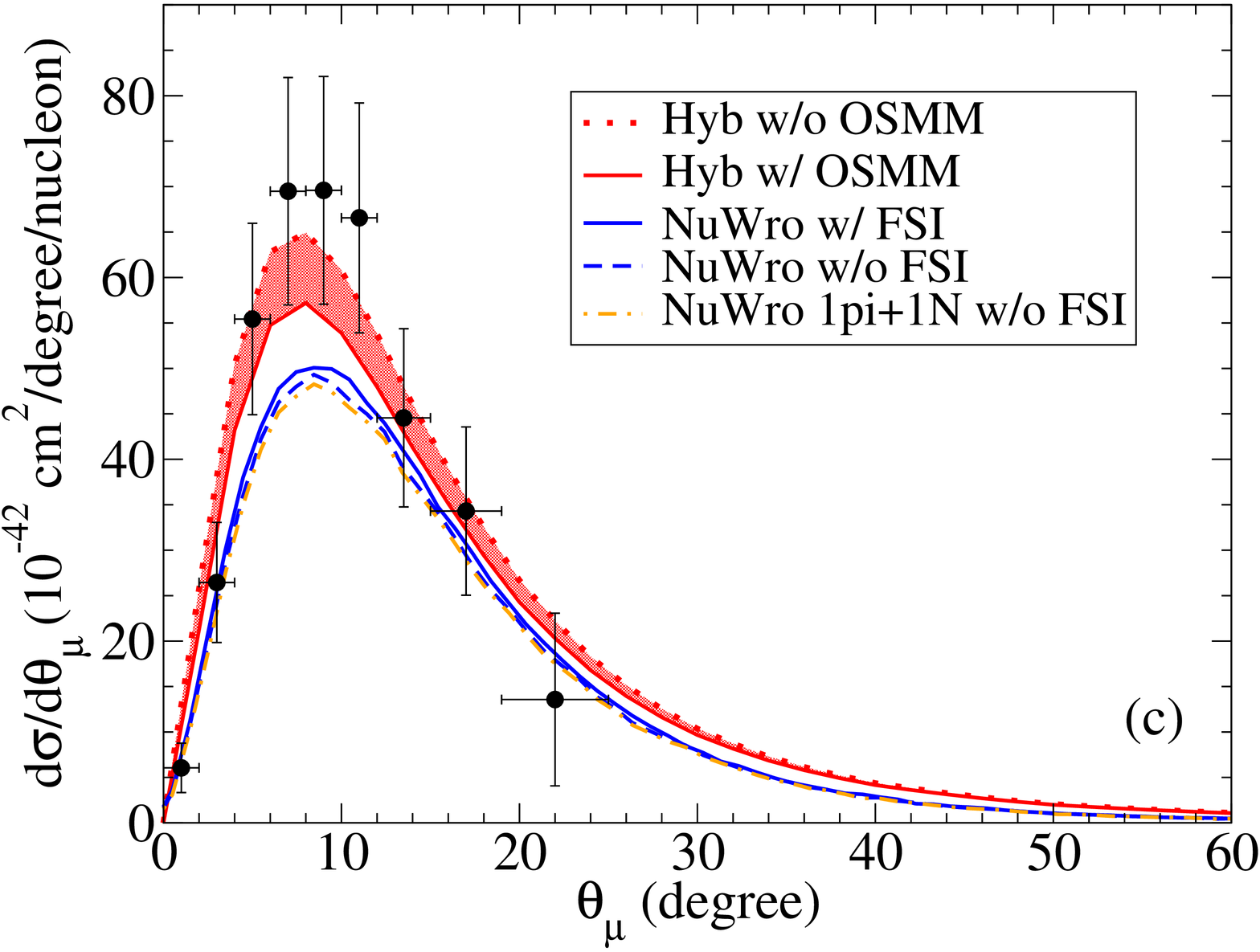}
      \includegraphics[width=.24\textwidth,angle=270]{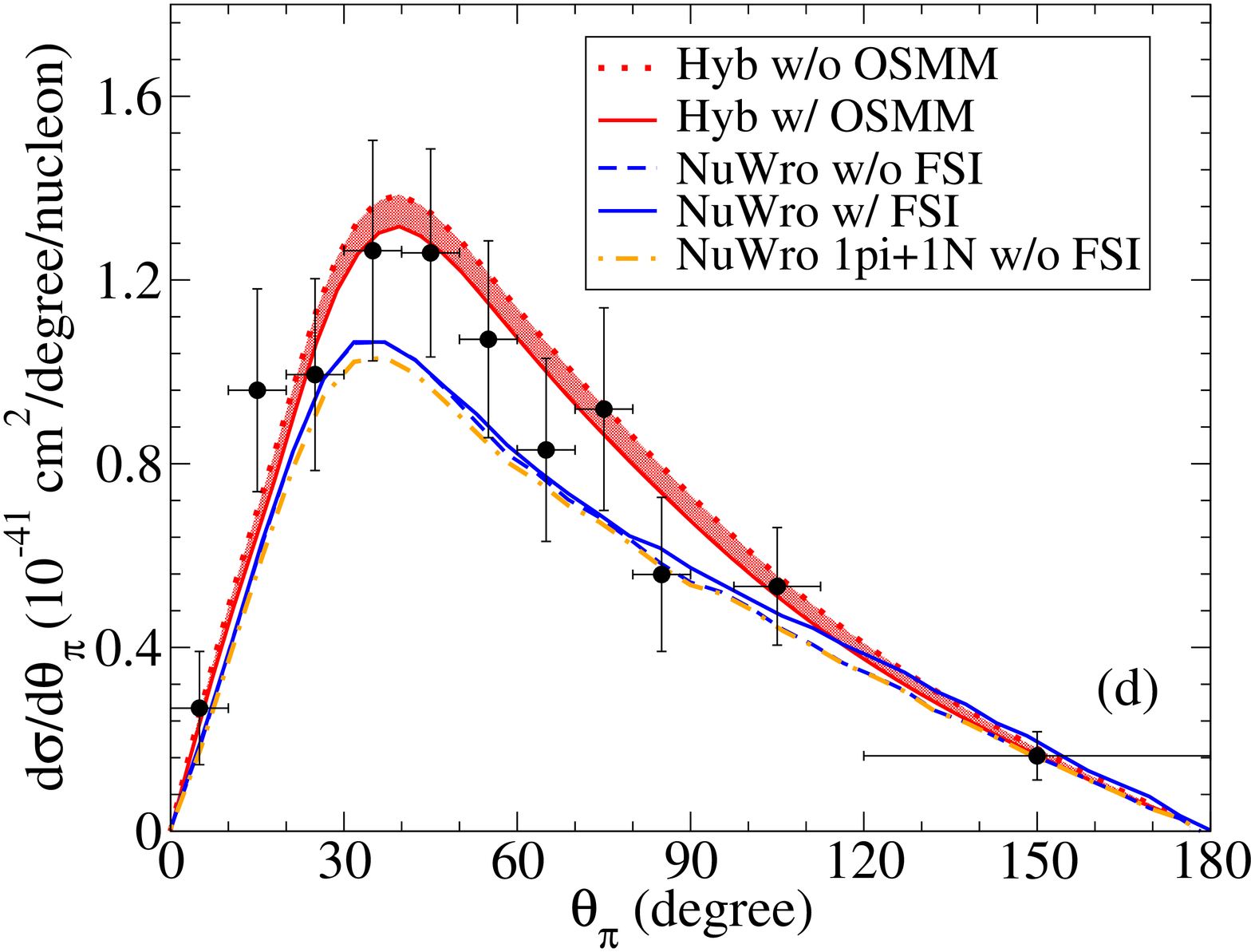}
  \caption{ Single-differential cross sections for the MINERvA $\bar\nu$CC $1\pi^0$ sample~\cite{MINERvACCpi16}. Labels as in Fig.~\ref{fig:MB-pi+}.}
  \label{fig:Min-nub-0pi} 
\end{figure*}

\begin{figure*}[htbp]
  \centering  
      \includegraphics[width=.22\textwidth,angle=270]{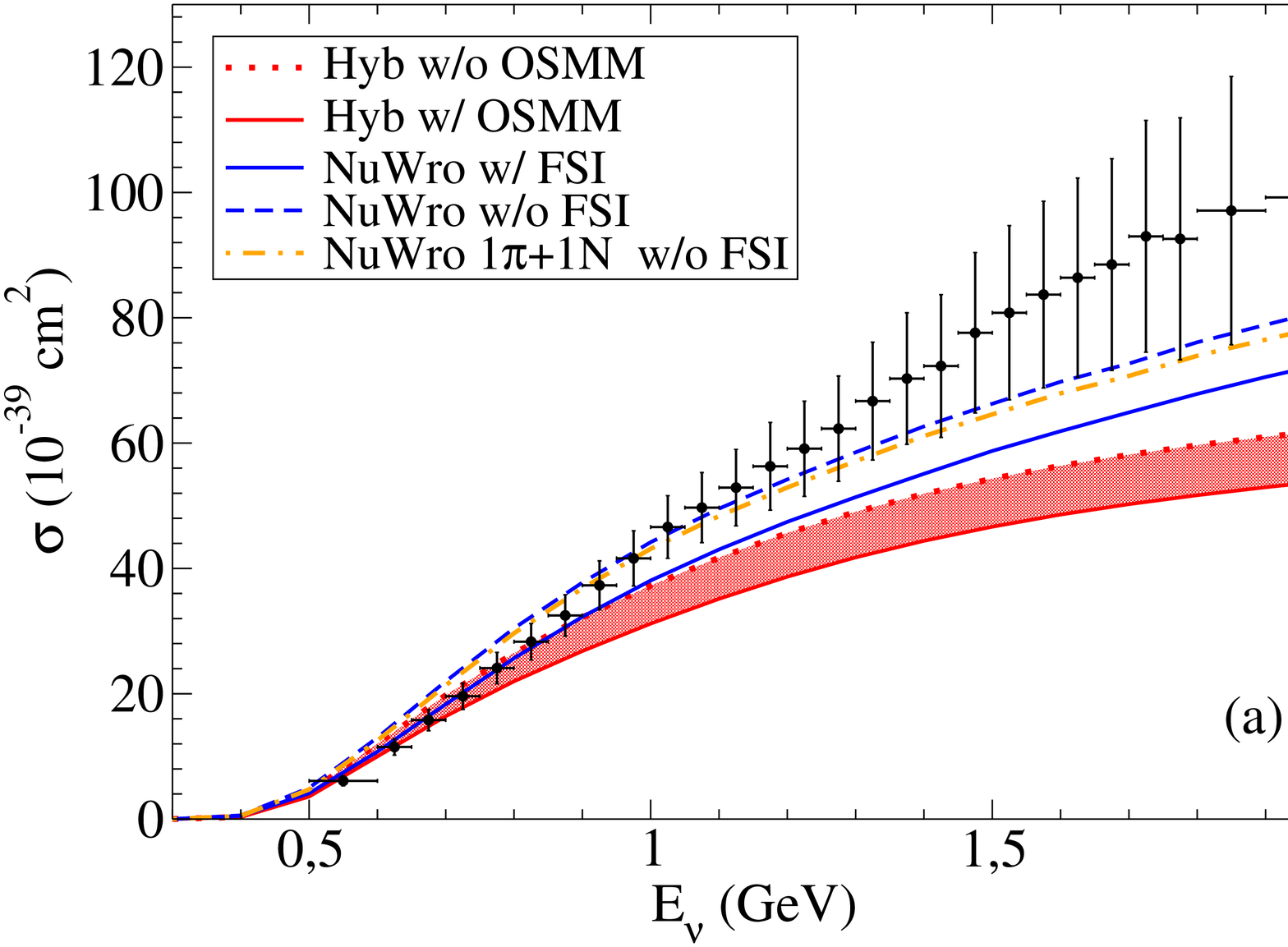}
      \includegraphics[width=.22\textwidth,angle=270]{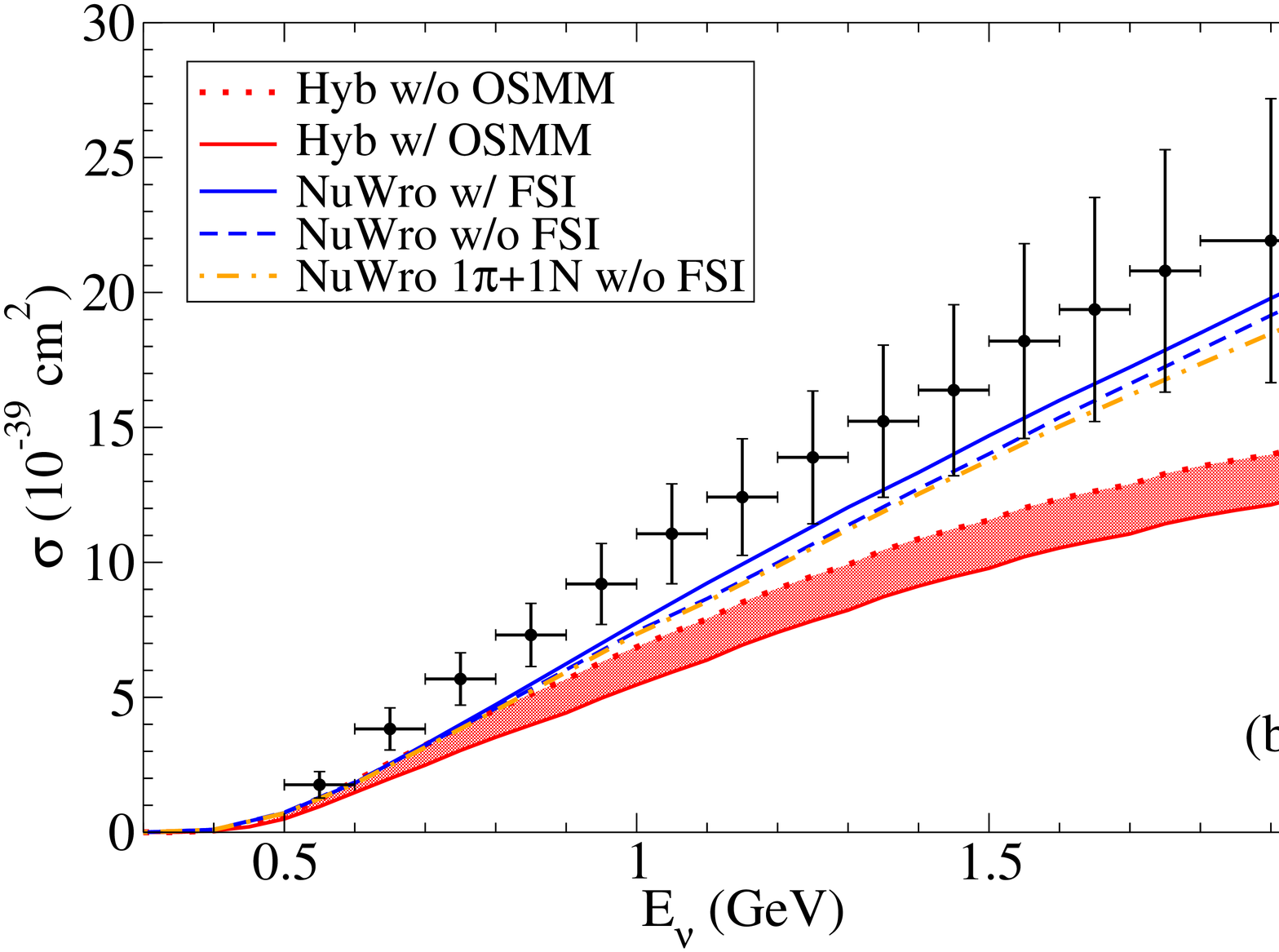}\\
      \includegraphics[width=.22\textwidth,angle=270]{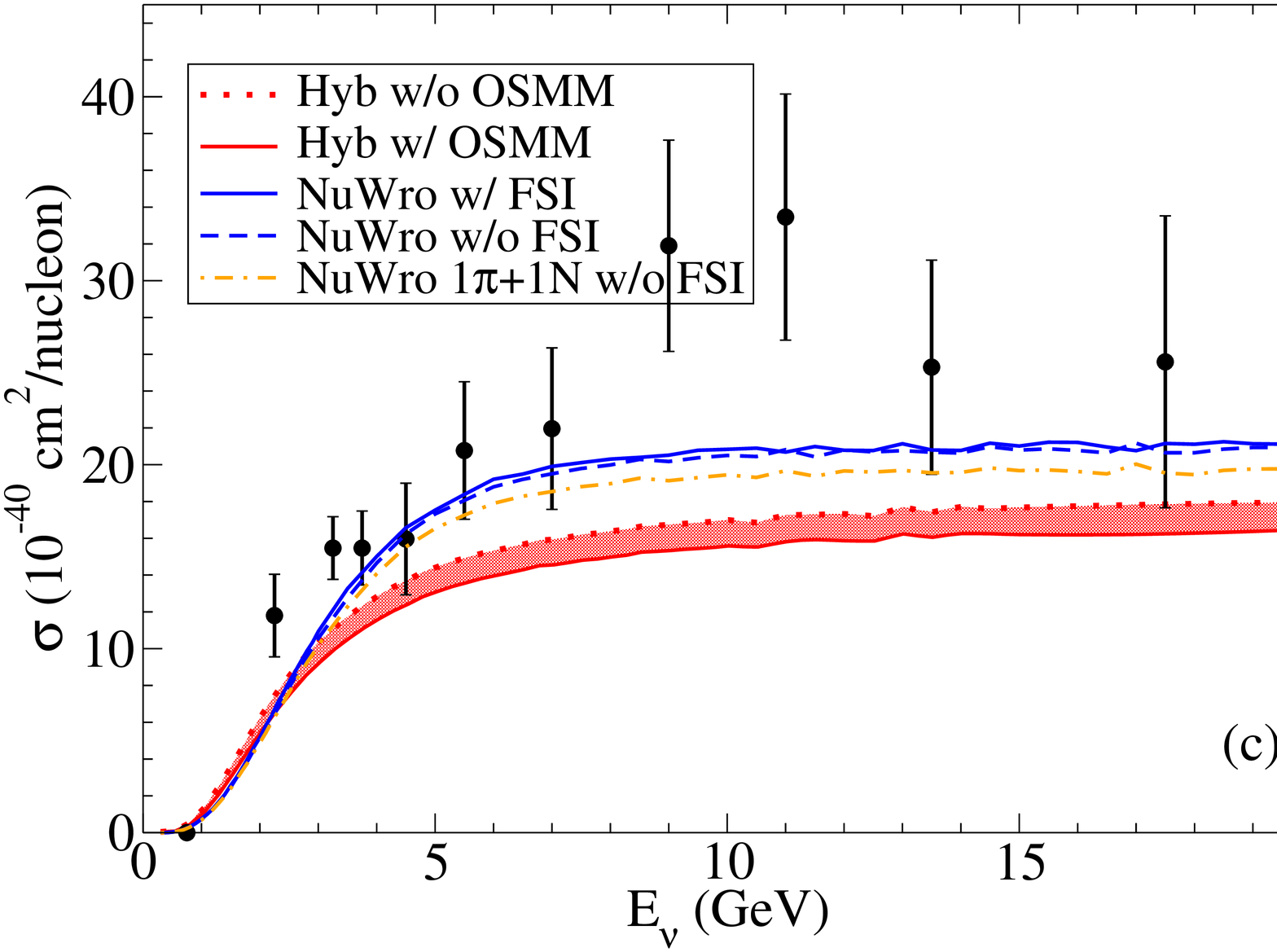}      
      \includegraphics[width=.22\textwidth,angle=270]{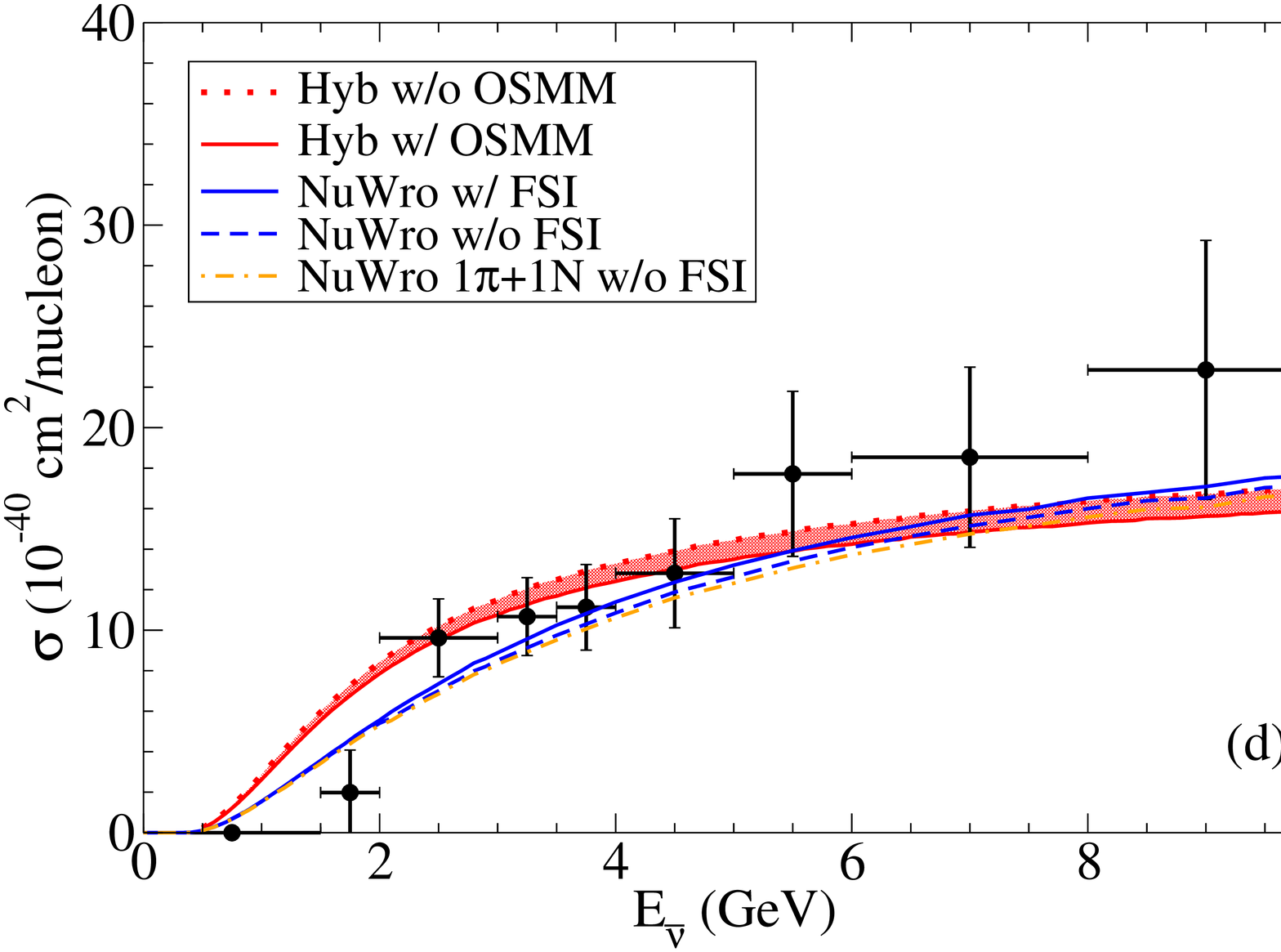}
  \caption{Total cross section for the reactions 
  (a) MiniBooNE $\nu$CC $1\pi^+$~\cite{MBCCpionC11}, 
  (b) MiniBooNE $\nu$CC $1\pi^0$~\cite{MiniBooNECCpi010}, 
  (c) MINERvA $\nu$CC $1\pi^0$~\cite{MINERvAnuCC17}, and 
  (d) MINERvA $\bar\nu$CC $1\pi^0$~\cite{MINERvACCpi16}. 
  Labels as in Fig.~\ref{fig:MB-pi+}. }
  \label{fig:tot} 
\end{figure*}

In Figs.~\ref{fig:MB-pi+}-\ref{fig:tot}, we present three different results from NuWro.
The blue-solid lines correspond to the case in which we use the same definition of the signal as in the experiment.
To study the effect of FSI, we show the results computed without FSI (blue-dashed lines).
Finally, the orange dash-dotted lines are the NuWro predictions without FSI and requiring that only one pion and one nucleon exit the nucleus. This constraint ensures that the DIS contribution is restricted to the single-pion production channel.
Therefore, these latter results correspond to the elementary SPP process predicted by NuWro, which can be compared with the hybrid-RPWIA results. 
Additionally, to make the comparison with NuWro more transparent, we have included the predictions of the hybrid-RPWIA model (with OSMM) when only the delta resonance and the background terms are considered, curves labeled as `Hyb w/ OSMM (Delta+Bgs)' in Figs.~\ref{fig:MB-pi+}-\ref{fig:tot}. This also allows us to study the effect of the higher mass resonances (P$_{11}$, S$_{11}$, and D$_{13}$) on the cross sections.

The single-differential cross sections for the $\nu$-induced 1$\pi^+$ production are shown in Fig.~\ref{fig:MB-pi+} (MiniBooNE) and Fig.~\ref{fig:Min-pi+} (MINERvA). 
NuWro predicts larger cross sections than the hybrid-RPWIA model. 
This was expected since the NuWro predictions for $\nu$-induced pion production off the nucleon are systematically larger than the ones from the hybrid model (see Fig.~19 in \cite{Gonzalez-Jimenez17}).
However, the shape of the cross sections computed with the hybrid-RPWIA model and NuWro (for the elementary reaction, dash-dotted line) are in good agreement. 
Regarding the FSI, charge exchange and pion absorption significantly reduce the amount of the $\pi^+$ that exit the nucleus. 
According to NuWro, this translates into a reduction of approximately 10-20\% in the magnitude of the cross section (dashed vs solid lines).
The effect of FSI is more relevant for the pion kinetic energy distribution, where one observes a redistribution of the strength that generates a pronounced peak in the region of small pion energies [Fig.~\ref{fig:MB-pi+}(b) and Fig.~\ref{fig:Min-pi+}(a)].
In general, FSI improve the agreement with MINERvA data but worsen the comparison with MiniBooNE, similarly to what was found in Refs.~\cite{Hernandez13,Lalakulich13b,Mosel17}.

In Figs.~\ref{fig:MB-pi0} and \ref{fig:Min-nu-0pi}, we show the single-differential cross sections for MiniBooNE and MINERvA $\nu$-induced $1\pi^0$ production, respectively.
The effect of FSI is very small because of the cancellation of two competing effects: the loss of $\pi^0$ due to charge exchange and absorption, and the creation of $\pi^0$ through the charge-exchange process $\pi^+ + n\rightarrow \pi^0 + p$. The latter mechanism tends to dominate due to the large amount of $\pi^+$ created in the reactions $p(\nu,\mu^-\pi^+)p$ and $n(\nu,\mu^-\pi^+)n$. 
The effect of FSI remains important in the pion kinetic energy distribution.
As in the $\nu$-induced 1$\pi^+$ production samples of Figs.~\ref{fig:MB-pi+} and \ref{fig:Min-pi+}, the NuWro predictions for the elementary reaction are, in general, larger than those from the hybrid-RPWIA model.
Both approaches, however, underpredict the data.

In Fig.~\ref{fig:Min-nub-0pi}, we present the results for the MINERvA $\bar\nu$-induced $1\pi^0$ production.
The effect of FSI is similar to that in Figs.~\ref{fig:MB-pi0} and \ref{fig:Min-nu-0pi}.
In this case, however, the hybrid-RPWIA predictions are above those of NuWro and in good agreement with data.
We have checked that, in general, NuWro predicts a larger asymmetry between the neutrino and antineutrino cross section than the hybrid model, which is a manifestation of a larger vector-axial interference response in NuWro. 
This is also evident by comparing Figs.~\ref{fig:tot}(c) and \ref{fig:tot}(d).

The total cross sections are presented in Fig.~\ref{fig:tot}.
The hybrid-RPWIA model underpredicts the MiniBooNE data [Fig.~\ref{fig:tot}(a) and \ref{fig:tot}(b)], except for the 1$\pi^+$ channel [Fig.~\ref{fig:tot}(a)] in the region $E_\nu<0.8$ GeV.
The disagreement is significant for neutrino energies above 1 GeV, especially in Fig.~\ref{fig:tot}(a), if one takes into account that the incorporation of FSI would reduce the cross section, moving theory further away from data. 
Since in the energy region of Figs.~\ref{fig:tot}(a) and (b), the hybrid model reproduces well the ANL-BNL deuterium data~\cite{Wilkinson14} (see Fig. 19 in Ref.~\cite{Gonzalez-Jimenez17}), we cannot readily provide an explanation for this disagreement.
NuWro also falls below the MiniBooNE data but shows a better agreement in the region $E_\nu>1$ GeV.

A similar situation is shown in Fig.~\ref{fig:tot}(c), where the hybrid-RPWIA model and NuWro underpredict the MINERvA $\nu$-induced 1$\pi^0$ production data.
Considerably better is the comparison between models and data for the MINERvA $\bar\nu$-induced $1\pi^0$ production sample [Fig.~\ref{fig:tot}(d)], in particular, in the region $2<E_{\bar\nu}<5$ GeV, which corresponds to the maximum of the antineutrino flux and where the experimental error bars are smaller.

The contribution of the higher mass resonances P$_{11}$, S$_{11}$, and D$_{13}$ in the $\pi^+$ samples (Figs.~\ref{fig:MB-pi+} and \ref{fig:Min-pi+}) is small due to the strong dominance of the $\Delta^{++}$ resonance, the low energy flux in the MiniBooNE sample and the cut $W_{rec}<1.4$ GeV in the MINERvA one. 
In the MiniBooNE $\pi^0$ sample (Fig.~\ref{fig:MB-pi0}), the effect is slightly larger than in the previous cases but still suppressed by the MiniBooNE flux.
In the MINERvA neutrino and antineutrino $\pi^0$ samples (Figs.~\ref{fig:Min-nu-0pi} and \ref{fig:Min-nub-0pi}), the less restrictive cut $W_{rec}<1.8$ GeV allows for larger contributions from the higher mass resonances, this results in an increase of the cross sections of more than 20\% in some cases.

\section{Conclusions}

The hybrid model for electroweak single-pion production off the nucleon, developed in Ref.~\cite{Gonzalez-Jimenez17}, has been extended here to the case of incoherent electroweak pion-production on the nucleus.

The pion-production mechanism includes the $s$- and $u$-channel diagrams for the $P_{33}(1232)$, $D_{13}(1520)$, $S_{11}(1535)$ and $P_{11}(1440)$ resonances, and the (first-order) background terms derived from the ChPT pion-nucleon Lagrangian~\cite{Hernandez07}. 
In our model, the high-energy behavior of the amplitude is given by a Regge approach~\cite{Gonzalez-Jimenez17}, this avoids the nonphysical behavior shown by the low-energy models when large invariant masses are explored.

Regarding the nuclear model, we used the relativistic impulse approximation to simplify the treatment of the nuclear current. The bound nucleons were described within the RMF model while the outgoing nucleon and pion are treated as plane waves, i.e., we did not consider FSI. 
We referred to the combination of the hybrid model and the RPWIA approach as the hybrid-RPWIA model.
This approach provides an estimate of the elementary reaction and can be used as a base for the implementation of FSI.

We have restricted our analysis to charge-current SPP,
though, the same model and code can be applied to electromagnetic and weak neutral-current interactions~\cite{Gonzalez-Jimenez17}. 
In particular, we have focused on the study of MiniBooNE and MINERvA kinematics, where the delta resonance is the main contribution. 
In Sec.~\ref{sec:MM}, the effect of the medium-modification of the delta-decay width was analyzed using the Oset and Salcedo prescription~\cite{Oset87}.
Due to the lack of a complete and consistent description of these medium modifications, we have considered them as an uncertainty, probably, the main one for the 1$\pi^+$ production channel under the kinematics explored here.

In Sec.~\ref{sec:lem-hyb}, we estimated the nonphysical strength that contaminates the cross sections when a low-energy model is used in cases where high-$W$ values are allowed in the physical phase space.
We conclude that for the MiniBooNE and MINERvA differential cross sections, the pathological high-$W$ behavior associated with the background terms is not significant [compare the solid and dash-dotted lines in Figs.~\ref{fig:hyb-lem}(c) and \ref{fig:hyb-lem}(d)]. In the case of MiniBooNE, this is due to the low-energy flux while in MINERvA this is due to the restriction $W_{exp}<1.4$ and $W_{exp}<1.8$ GeV in the 1$\pi^+$ and 1$\pi^0$ samples, respectively. 
On the contrary, we observed that it is important to regularize the amplitudes of the resonances in all cases (compare the dashed and dash-dotted lines in Fig.~\ref{fig:hyb-lem}). 
The nonphysical contributions from the background terms are relevant in the case of MiniBooNE when studying the total cross section. In this case, the pathologic contributions appear at $E_\nu\approx$1.2 GeV and keep growing for increasing energies [compare the dash-dotted and solid lines in Fig.~\ref{fig:hyb-lem}(a)]. Notice that, the MiniBooNE sample does not contain any restriction on the invariant mass. 
 
In Sec.~\ref{sec:HybridvsNuWro}, we compared the hybrid-RPWIA model and NuWro predictions with the MiniBooNE and MINERvA data. 
The goal of comparing with NuWro was twofold. 
First, this allowed us to estimate the effect of FSI by analyzing the NuWro results with and without FSI. 
Second, by turning off FSI and restricting the definition of the signal to only one pion and one nucleon in the final state (`NuWro 1$\pi$+1N w/o FSI'), we were able to compare the NuWro predictions of the elementary SPP reaction with those from the hybrid-RPWIA model. 
Although the two approaches are completely different in both the description of the elementary vertex and the nuclear dynamics, ideally, one would expect the two predictions (`NuWro 1$\pi$+1N w/o FSI' and hybrid-RPWIA) to match each other. 
The results presented here, however, show that we are far from this {\it ideal} case.
More investigation and comparison between models, on the elementary reaction and the FSI mechanisms, will be needed before more definite conclusions can be made.

Finally, using the hybrid-RPWIA model, we have shown that the higher mass resonances $D_{13}(1520)$, $S_{11}(1535)$ and $P_{11}(1440)$ have a relatively small effect on the MINERvA $\pi^+$ and MiniBooNE samples.
On the contrary, for the MINERvA neutrino and antineutrino $\pi^0$ samples, the contribution from these resonances is important, increasing the cross sections by more than 20\% at some kinematics.

The natural continuation of this project is the implementation of the elastic distortion of the outgoing nucleon and pion wave functions. 
The inelastic FSI can be treated by implementing the model in an MC event generator.

\section*{Acknowledgements}

This work was supported by the Interuniversity Attraction Poles Programme initiated by the Belgian Science Policy Office (BriX network P7/12) and the Research Foundation Flanders (FWO-Flanders), and partially by the Special Research Fund, Ghent University. 
The computational resources (Stevin Supercomputer Infrastructure) and services used in this work were provided by Ghent University, the Hercules Foundation and the Flemish Government.
K.N. was partially supported by the Polish National Science Center (NCN), under Opus Grant No. 2016/21/B/ST2/01092, as well as by the Institute of Theoretical Physics, University of Wroc{\l}aw Grant No. 0420/2545/17.
We want to thank J.M. Ud\'ias and J.A. Caballero for useful discussions about the Relativistic Mean-Field model, and J. Sobczyk and T. Golan for their help concerning the NuWro generator.

\bibliographystyle{apsrev4-1}
\bibliography{bibliography}

\end{document}